\documentclass[12pt,abstracton]{scrartcl}

\usepackage[T5,T1]{fontenc}  

\usepackage{microtype}
\usepackage{soul}
\usepackage{color}
\usepackage{paralist}
\usepackage[ruled]{algorithm2e}

\usepackage{amsmath}%
\usepackage{amsfonts}%
\usepackage{amsthm}
\usepackage{amssymb}%
\usepackage{graphicx}%
\usepackage{booktabs}
\usepackage{tikz}
\usepackage{pgfkeys}
\usepackage{pgfplots}
\usepackage{algorithmic}%
\usepackage{url}
\usepackage{hyperref}
\usepackage{subfigure}
\usepackage{multirow}
\usepackage[normalem]{ulem}

\algsetup{linenosize=\small}

\usetikzlibrary{%
  arrows,%
  calc,%
  fit,%
  patterns,%
  plotmarks,%
  shapes.geometric,%
  shapes.misc,%
  shapes.symbols,%
  shapes.arrows,%
  shapes.callouts,%
  backgrounds,%
  chains,%
  topaths,%
  trees,%
  petri,%
  mindmap,%
  matrix,%
  folding,%
  fadings,%
  through,%
  positioning,%
  scopes,%
  decorations.fractals,%
  decorations.shapes,%
  decorations.text,%
  decorations.pathmorphing,%
  decorations.pathreplacing,%
  decorations.footprints,%
  decorations.markings,%
  shadows}

\pgfplotsset{every axis/.append style={font=\large, line width=1pt, tick style={line width=2pt}}}
\pgfplotsset{compat=1.3}
\tikzset{every mark/.append style={scale=1.5}}

\tikzset{line/.style=}
\tikzset{
        hatch distance/.store in=\hatchdistance,
        hatch distance=2pt,
        hatch thickness/.store in=\hatchthickness,
        hatch thickness=1pt
    }

\newtheorem{theorem}{Theorem}

\newtheorem{example}{Example}

\newtheorem{property}{Property}

\numberwithin{equation}{section}

\newcommand{\eat}[1]{{}}

\begin{document}

\title{Efficient Network-aware Search in Online Social Bookmarking Applications} 


\author{Silviu~Maniu ~ ~ ~ ~ Bogdan Cautis\\
        T\'{e}l\'{e}com ParisTech, Paris, France\\
        first.last@telecom-paristech.fr
        }
\date{}
\maketitle

\begin{abstract}We consider in this paper top-k query answering in social tagging (or bookmarking) applications. This problem requires a significant departure from existing, socially agnostic techniques.  In a network-aware context,  one can (and should) exploit the social links, which  can indicate how  users relate to the seeker  and  how much weight their tagging actions should have in the result build-up.  We propose an algorithm that has the potential to scale to current applications. While the problem has already been considered in previous literature, this was done either under strong simplifying assumptions or under choices that cannot scale to even moderate-size real-world applications. We first revisit a key aspect of the problem, which is accessing the closest or most relevant users for a given seeker. We describe how this can be done on the fly (without any pre-computations) for several possible choices - arguably the most natural ones - of proximity computation in a user network.   Based on this, our top-k algorithm is sound and complete,  while addressing the  applicability issues of the existing ones. Moreover, it performs significantly better and, importantly, it is  instance optimal in the case when the search relies  exclusively on the social weight of tagging actions.    To further reduce response times, we then consider directions for efficiency by approximation. Extensive experiments on real world data show that our techniques can drastically improve the response time, without sacrificing precision.
\eat{
We consider in this paper top-k query answering in social tagging systems, also known as folksonomies. This problem requires a significant departure from existing, socially agnostic techniques.  In a network-aware context,  one can (and should) exploit the social links, which  can indicate  how  users relate to the seeker  and  how much weight their tagging actions should have in the result build-up.  We propose an algorithm that has the potential to scale to current applications. While the problem has already been considered in previous literature, this was done either under strong simplifying assumptions or under choices that cannot scale to even moderate-size real world applications. We first consider a key aspect of the problem, which is accessing the closest or most relevant users for a given seeker. We describe how this can be done on the fly (without any pre-computations) for several possible choices - arguably the most natural ones - of proximity computation in a user network. 
 Based on this, our top-k algorithm is sound and complete, and exhibits the same behavior as the one from existing literature, while addressing its scalability issues. To further reduce response times, we also analyze in more depth this behavior a Del.icio.us dataset and identify  promising directions for efficiency by approximation.
}
\end{abstract}



\section{Introduction}\label{sec:introduction}
Unprecedented volumes of data are now at  everyone's fingertips on the World Wide Web. The ability to query them efficiently and effectively, by fast retrieval and ranking algorithms, has largely contributed to the rapid growth  of the Web, making it simply  irreplaceable in our every day life.

 A new dynamics to this development has been recently brought by the \emph{social Web},  applications that are centered around users, their relationships and their data. Indeed, user-generated content is becoming a significant and highly qualitative portion of the Web. To illustrate, the most visited Web site today is a social one.
 This calls for adapted,  efficient retrieval techniques,  which can go beyond a  classic Web search paradigm  where  data is decoupled from the users querying it.

 An important  class of social applications are the \emph{collaborative tagging applications}, also known as  \emph{social bookmarking applications},  with popular examples including Del.icio.us, StumbleUpon or Flickr. Their general setting is the following:
\begin{itemize}
\item users form a \emph{social network}, which may reflect proximity, similarity, friendship, closeness, etc,
\item items from a public pool of items (e.g., document, URLs, photos, etc) are \emph{tagged} by users with keywords, for purposes such as description and classification, or to facilitate later retrieval,
\item users \emph{search} for items having certain keywords (i.e., tags)  or they are \emph{recommended} items, e.g.,  based on proximity at the level of tags.
\end{itemize}
Collaborative tagging, and social applications in general,   can offer an entirely new perspective to how one searches and accesses information.  The main reason for this is that  users can (and often do) play a role at both ends of the information flow,  as producers and also as  seekers of information. Consequently,  finding the most relevant items that are tagged by some keywords should be done in a \emph{network-aware} manner. In particular,  items that are tagged by  users who are ``closer''  to the seeker -- where the term closer depends on model assumptions that will be clarified shortly --   should be given more weight  than  items that are  tagged by more distant users.

We consider in this paper the problem of top-$k$ retrieval in collaborative tagging systems. We  investigate it with a focus on efficiency, targeting techniques that have the potential to scale to current applications on the Web\footnote{The most popular ones  have user bases of the order of millions and huge repositories of data;  today's most accessed  social Web application, which also provides tagging and searching functionalities, has more than half a billion registered users.}, in an online context where the social network,  the tagging data and even the seekers' search ingredients can change at any moment. 
In this context, a  key sub-problem for top-$k$ retrieval  that we need to address is computing scores of top-$k$ candidates by iterating not only through the most relevant items  with respect to the query, but also (or mostly) by looking at the closest users and their tagged items.
  
We associate with the notion of social network a rather general interpretation, as a user graph whose edges are labeled by \emph{social scores}, which give a measure of the  proximity or similarity  between two  users. These are  then exploitable in searches, as they say how much weight one's tagging actions should have in the result build-up. For example, even for tagging applications where an explicit social network does not exist or is not exploitable, one may use the tagging history to build a network based on  similarity 
in tagging and items of interest.
While we focus mainly on bookmarking applications, we believe that these represent a good abstraction for other types of social applications, to which our techniques could directly apply.



\begin{figure}
	\label{fig:network}
    \centering
		\includegraphics[scale=0.50]{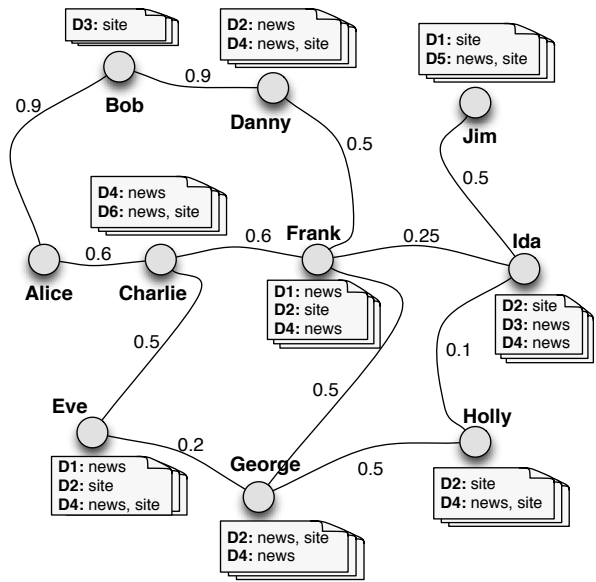}
\caption{A collaborative tagging scenario and its social network.}
\end{figure}
\begin{example} 
{\emph Consider the collaborative tagging configuration of Figure~\ref{fig:network}. Users have  associated lists of tagged documents and  they are  interconnected by social links. Each link is labeled by its (social) score, assumed to be in the $[0,1]$ interval. Let us consider user $Alice$ in the role of the seeker. The user graph is not complete, as the figure shows,  and only two users have an explicit social score with respect to $Alice$. For the remaining ones, $Danny, \dots, Jim$, only an implicit social score  could be computed from the existing links if a precise measure of their relevance with respect to $Alice$'s queries is necessary in the top-$k$ retrieval.

Let us assume that $Alice$ looks for the top two documents that are  tagged with both $news$ and $site$. Looking at $Alice$'s immediate neighbors and their respective documents,  intuitively,  $D3$ should have a higher score than $D4$,  since the former is tagged by a  more relevant user ($Bob$, having the maximal social score relative to $Alice$).  If we expand the search to the entire graph,  the score of $D4$ may however benefit from the fact that other users, such as $Eve$ or even $Holly$, also tagged it with $news$ or $site$.  Furthermore, documents such as $D2$ and $D1$ may also be relevant for the top-$2$ result, even though they were tagged  only by users who are indirectly linked to $Alice$.

Under certain assumptions to be clarified shortly, the top-$2$ documents for $Alice$'s query will be, in descending score order, $D4$ and $D2$. The rest of the paper will present the underlying model and algorithms that allow us to build this answer.
}
\label{ex:setting}
\end{example}

\textbf{Main related work.} Classic top-$k$ retrieval algorithms, such as Fagin's threshold algorithm~\cite{Fagin01} and the no random access (NRA)  algorithm,  rely on precomputed inverted-index lists with exact scores for each query term (in our setting, a term is a tag).   
  Revisiting the setting in Figure~\ref{fig:network}, we would have two per-tag inverted lists $IL(news)=\{D4:7,  D2:2, D1:2, D3:1, D6:1, D5:1\}$ and $IL(site)=\{D2:5, D4:2, D3:1, D6:1, D1:1, D5:1\}$,  which give the number of times a document has been tagged with the given tag.
 
 When user proximity is an additional ingredient in the top-$k$ retrieval process,   a direct network-aware adaptation of the threshold algorithm and variants  would  need  precomputed in\-verted-index lists for each user-tag pair. 
For instance, if we interpret explicit links in the user graph as friendship, ignoring the link scores, and only tagging by direct friends matters,  $Alice$'s lists would be $IL_{Alice}( news)=\{D4:1,D6:1\}$ and $IL_{Alice}(site)=\{D3:1,D6:1\}$. Other $18$ such lists would be required   
  and,  clearly, this would have  prohibitive space  and computing costs in a real-world setting.     
  Amer-Yahia et al.~\cite{Amer-Yahia08} is the first to address this issue, considering the problem of network-aware search in collaborative tagging sites, though by a simplified flavor.  The authors consider  an  extension to classic  top-$k$ retrieval in which user proximity is seen as   a binary function (0-1 proximity): only a subset of the users in the network are selected and \eat{users who are directly connected to the seeker} can influence the top-$k$  result.  This introduces two strong simplifying restrictions: (i) only documents tagged by the selected users should be relevant in the search, 
  and (ii) all the \eat{friends of a given seeker} users thus selected are equally important. 
 The base solution of~\cite{Amer-Yahia08} is to  keep for each tag-item pair, instead of the detailed lists per user-tag pair, only an upper-bound value on the number of taggers.\eat{: the maximal number of taggers from any user's neighborhood.}  
For instance, the upper-bound for $(news, D4)$ would be  $2$, since for any user there are at most two neighbors who tagged $D4$ with $news$. This is called  the \textsc{Global Upper-Bound} strategy. A more refined version, which trades space for efficiency, keeps such upper-bound values within \emph{clusters }of users,  instead of the network as a whole.
\eat{
The main drawback of~\cite{Amer-Yahia08}'s approach is that it cannot  apply under a more general social search interpretation, in which a tagger's importance for the top-$k$ result depends on how strongly she is related to the seeker, \emph{without necessarily being a direct friend thereof.}
}
   
Only  in Schenkel et al.~\cite{Schenkel08}, the network-aware retrieval problem for collaborative tagging is  considered under a general interpretation, the one we also adopt in this paper. It considers that even users  who are only indirectly connected to the seeker can be relevant for the top-$k$ result. 
Their 
 \textsc{ContextMerge} algorithm follows the intuition that the users closest to the seeker will contribute more to the score of an item, thus maximizing the chance that the item will remain in the final top-$k$. The authors describe a hybrid approach in which, at each step, the algorithm chooses either to look at the documents tagged by the closest unseen user or at the tag-document inverted lists (a seeker agnostic choice).  In order to obtain the next (unseen) closest user at any given step, the algorithm  \emph{precomputes} in advance the proximity value for all possible pairs of users. These values are then stored in ranked lists  (one list per user), and  a simple pointer increment allows  to obtain the next relevant user.
 \begin{example} Consider the network of  Fig.~\ref{fig:network}. With respect to seeker $Alice$, the list of users ranked by proximity would be $\{Bob:0.9, Danny:0.81, Charlie:0.6, Frank:0.4, Eve:0.3, George:0.2, Holly:0.1,  Ida:0.1, Jim:0.05\}$, 
 with proximity between two users  built as the \emph{maximal product of scores over paths linking them} (formalized in Section~\ref{sec:dijkstra}).
 \label{ex:mulprox}
\end{example}
The main drawbacks of~\cite{Schenkel08} are scalability and applicability. Clearly, precomputing a weighted  transitive closure  over the entire network  has a high cost in terms of space and computation in even moderate-size social networks. More importantly, keeping these proximity lists up to date when they reflect tagging similarity\footnote{Tagging similarity may indeed be a more pertinent proximity measure than friendship for top-$k$ search in bookmarking applications.} (as advocated in~\cite{Schenkel08}), would simply be unfeasible in real-world settings,  which are highly dynamic. (We revisit these considerations in Section ~\ref{sec:performance}.)  


 \textbf{Main contributions.} We propose an algorithm  for top-$k$ answering in collaborative tagging, which has the potential to scale to current applications and beyond, in an online context where network changes  and tagging actions are frequent. For this algorithm, we first address a  key aspect: accessing efficiently the closest users  for a given seeker. We describe how this can be done \emph{on the fly} (without any pre-computations) for a large family of functions for proximity computation in a social network, including the most natural ones (and the one assumed in~\cite{Schenkel08}).  The interest in doing this is threefold:
\begin{itemize}
    \item we can support full scoring personalization, where each user issuing queries can define her own way to rank items, through parameters and score function choices,
    \item we can iterate over the relevant users  in more efficient manner,  since a typical network can easily fit in main-memory; this can spare the potentially huge disk volumes  required by~\cite{Schenkel08}'s algorithm (see Section~\ref{sec:performance}), while  also having the potential to run faster. 
    \item social link updates are no longer an issue; in particular, when the social network depends on the tagging history, we can keep it up-to-date  and, by it, all the proximity values at any given moment,  with  little overhead.
\end{itemize}

Based on this, our top-$k$ algorithm \textsc{TOPKS}\eat{\footnote{Algorithm name obfuscated for double-blind reviewing.}} is sound and complete. We show that, when the search relies exclusively on the social weight of tagging actions,  it is  \emph{instance optimal} in a large and important class of algorithms.  Extensive experiments on real world data show that our algorithm performs significantly better than existing techniques, with up to 50\% improvement (see Section~\ref{sec:results}).

For further efficiency, we then consider  directions for approximate results. Our approaches present the advantages of negligible memory consumption (they rely on concise statistics about the user network) and reduced computation overhead. Moreover, these statistics  can be maintained up to date with limited effort, even when the social network is built based on tagging history. Experiments show that  approximate search techniques can drastically improve the response time, reaching around 25\% of the running time of the exact approach, without sacrificing precision. 

\emph{
The main focus of our work is on the social aspects of top-$k$ retrieval in collaborative tagging applications, and our techniques are designed to perform best in settings where tagging actions are mostly (if not exclusively) viewed through the lens of social relevance. 
}

\textbf{Outline.} The rest of the paper is organized as follows. In Section~\ref{sec:setting} we formalize the top-$k$ retrieval problem in collaborative tagging applications. We describe a key aspect of our approach, the on-the-fly computation of proximity in Section~\ref{sec:dijkstra}. We then describe our top-$k$ algorithm, first in an exclusively social form, in Section~\ref{sec:algorithm0}, and show it is instance optimal in Section~\ref{sec:optim}. The general  algorithm is then presented in Section~\ref{sec:alg_general}. Two approaches for  improving efficiency by approximation are given in Section~\ref{sec:approx}. We discuss applicability and scalability issues in Section~\ref{sec:performance}. Experimental results are presented in Section~\ref{sec:results}.
We overview the related work in Section~\ref{sec:related}.
We  discuss future work and we conclude in Section~\ref{sec:future}.


\section{General Setting}\label{sec:setting}
\noindent We consider a social setting in which we have a set of items (could be text documents, URLs, photos, etc) ${\cal I}=\{i_{1},\dots,i_{m}\}$, each tagged with one or more  distinctive tags from a dictionary of tags  ${\cal T}=\{t_{1},t_{2},\dots,t_{l}\}$ by one or more  users from ${\cal U}=\{u_1, \dots, u_n\}$. We assume that users form an undirected weighted graph $G=({\cal U}, E, \sigma)$ called the \emph{social network}.  In  $G$, nodes represent  users and $\sigma$ is a function that associates to each edge $e=(u_1, u_2)$ a value in $(0,1]$, called  \emph{the proximity} (or social) score  between $u_1$ and $u_2$.  

\emph{Given a seeker user $s$, a keyword  query $Q=(t_1,...,t_r)$ (a set of $r$ distinct tags) and an integer value $k$, the top-$k$ retrieval problem is to compute the (possibly ranked) list of the $k$ items  having the highest scores with respect to the seeker and query.  
}


We describe next the score model for this problem. 

Extending  the model for social tagging systems presented in~\cite{Amer-Yahia08}, we also assume the following two relations for tags:
\begin{itemize}
	\item \textbf{tagging}: $Tagged(v,i,t)$: says that a user $v$ tagged the  item $i$ with tag $t$, 
	\item \textbf{tag proximity}: $SimTag(t_1, t_2, \lambda)$: says that tags $t_1$ and $t_2$ are similar, with similarity value $\lambda \in (0,1)$. 
\end{itemize}
We assume that a user can  tag a given item with a given tag at most once.   We first model for a  user, item and tag triple $(s, i, t)$ the \emph{score} of item $i$ for the given seeker $s$ and tag $t$. This is denoted $score(i~|~ s, t)$. Generally,
\begin{equation}
  score(i~|~s,t)=h(fr(i~|~s,t))
  \label{equ:score}
\end{equation}
where $fr(i~|~ s, t)$ is the \emph{overall term frequency}  of item $i$ for seeker $s$ and tag $t$, and $h$ is a positive monotone function.

The overall term frequency function $fr(i~|~ s, t)$ is defined as a combination of a network-dependent component and a document-dependent one, as follows:
\begin{equation}
	fr(i~|~ s, t)=\alpha \times  tf(t, i) +( 1-\alpha) \times sf(i~|~ s, t).
	\label{equ:socfreq}
\end{equation}
The former component, $tf(t, i)$, is the term frequency of $t$ in $i$, i.e., the number of times $i$ was tagged with $t$. The latter component stands for social frequency, a measure that depends on the seeker.\footnote{The linear combination of  Eq.~(\ref{equ:socfreq}) is one that is widely used when a local retrieval score and a global one are to be combined, e.g., in spatial search~\cite{Cao10} or in social search~\cite{Schenkel08}.  However, any monotone combination of the two score components can be used in these approaches, as in ours.}

If we consider that each user brings her own weight (proximity) to the score of an item, we can define the measure of social frequency as follows:
\begin{equation}
  sf(i~|~ s, t)=\sum_{v\in \lbrace v ~|~Tagged(v,i,t))\rbrace} \sigma(s,v).
  \label{equ:ScoresWeightSum}
\end{equation}
Then, given a query $Q$ as a set of tags $(t_1, \dots, t_r)$, the overall score of $i$ for seeker $s$ and query $Q$,   $$score(i~|~ s, Q)=g(score(i~|~ s, t_1), \dots ,score(i~|~ s, t_r)),$$ is   obtained using a monotone aggregate function $g$ over the  individual scores for each tag. In this paper,  the aggregation function $g$ is assumed to be a summation,  $g=\sum_{t_j\in Q}score(i~|~s,t_j)$.

\textbf{Extended proximity.} The above scoring model takes into account only the neighborhood  of the seeker (the users directly connected to her). But this can be extended to deal also with users that are indirectly connected to the seeker, following a natural interpretation that user links (e.g., similarity or trust) are (at least to some extent) transitive.   We denote by $\sigma^+$ an \emph{extended proximity}, which is to be computable from $\sigma$  for any pair of users connected by  a path in the network. Now,  $\sigma^+$ can replace $\sigma$ in the definition of social frequency we consider before  (Eq.~(\ref{equ:ScoresWeightSum})), yielding an overall item scoring scheme that depends on the entire network instead of only the seeker's vicinity.  We discuss shortly possible alternatives for  $\sigma^+$ by means of aggregating $\sigma$ values along paths in the graph.   In the rest of this paper, when we talk about proximity  we refer to the extended one. 

For a given seeker $u$, by her \emph{proximity vector} we denote the list of users with non-zero proximity with respect to $u$, ordered in descending order of these proximity values.    

\textbf{Remark 1.} In Eq.~(\ref{equ:socfreq}), the $\alpha$ parameter allows to tune the relative importance of the social component with respect to classic term frequency. When $\alpha$ is valued $1$, the score  becomes network-independent. On the other hand, when $\alpha$ is valued $0$ the score depends exclusively on the social network. 

\textbf{Remark 2.} Note that a network in which all the user pairs have a proximity score of $1$  amounts to the classical document retrieval setting (i.e., the result is independent of the user asking the query).

\textbf{Remark 3.} Tag similarity can be integrated into Eq.~(\ref{equ:ScoresWeightSum}), e.g., by setting a threshold $\tau$ s.t. if  $SimTag(t,t',\lambda)$,  with $\lambda$   above $\tau$, and $Tagged(v,i,t')$, we also add $\sigma(u,v)$ to  $sf(i~|~ u, t)$. For the sake of simplicity this is ignored in this paper, but remains an integral part of the model.

\textbf{Remark 4.}  Note that queries are not assumed to use only tags from ${\cal T}$. For any tag outside this dictionary, items will obviously have a score of $0$.
\subsection{Computing $\boldsymbol{\sigma^+}$}\label{sec:dijkstra}
We describe in this section a key aspect of our  algorithm for top-$k$ search, namely on-the-fly computation of proximity values with respect to a seeker $s$. The issue here is to facilitate at any given step the retrieval of the most relevant unseen user $u$ in the network, along with her proximity value $\sigma^+(s,u)$. This user will have the potential to contribute the most to the partial scores of items that are still candidates for the top-$k$ result, by Eq.~(\ref{equ:score}) and (\ref{equ:ScoresWeightSum}).
 
We start by discussing possible candidates for $\sigma^+$, arguably the most natural ones, drawing inspiration from studies in the area of trust propagation  for belief statements. We then give a  wider characterization for the family of possible functions for proximity computation, to which these candidates belong.



\emph{Candidate 1($f_{mul}$).} Experiments on trust propagation in the Epinions network  (for computing a final belief in a statement)~\cite{Richardson03} or in P2P networks show that  (i) multiplying the weights on a given path between $u$ and $v$, and  (ii) choosing the maximum value over all the possible paths,  gives the best results (measured in  terms of precision and recall) for predicting beliefs.  We can integrate this into our scenario, by assuming that  belief refers to \textit{tagging with a tag $t$}. We thus  aggregate the weights on  a  path  $p=(u_1, \dots, u_l)$ (with a slight abuse of notation) as 
$$\sigma^+ (p)= \prod_i \sigma(u_{i},u_{i+1}).$$

For seeker $Alice$ in our running example, we gave in the previous section (Example~\ref{ex:mulprox}) the proximity values and the ordering of the network under this candidate for $\sigma^+$. 

\emph{Candidate 2($f_{min}$).} A possible drawback of Candidate $1$ for proximity aggregation is that values may decrease quite rapidly.  A  $\sigma^+$ function that avoids this could be obtained by replacing multiplication over a path with minimal, as follows:    
$$\sigma^+ (p)= \min_i\{ \sigma(u_{i},u_{i+1})\}.$$ 

Under this  $\sigma^+$ candidate, the values with respect to seeker $Alice$ would be the following: $\{Bob:0.9, Danny:0.9, Charlie:0.6, Frank:0.6, Eve:0.5, George:0.5, Harry:0.5, Ida:0.25, Jim:0.25\}.$

\emph{Candidate 3($f_{pow}$).} Another possible   definition for $\sigma^+$  we consider 
relies on an aggregation that penalizes long paths, i.e., distant users, in a controllable way, as follows: 
$$\sigma^+ (p)= \lambda^ {-\sum_i\frac{1}{\sigma(u_{i},u_{i+1})}}.$$
where $\lambda \geq 1$  can be seen as a ``drop parameter''; the greater its  value the more rapid the decrease of proximity values. 
Under this candidate for $\sigma^+$, for $\lambda=2$, the rounded values w.r.t seeker $Alice$ would be 
$\{Bob:0.46, Charlie:0.31, Danny:0.21, Eve:0.077, Frank:0.0525, George:0.013, Ida:0.003, Harry:0.003, Jim:0.0007\}.$\\

The key common feature of the candidate functions previously discussed is that they are monotonically decreasing over any path they are applied to, when $\sigma$ draws values from the interval  $[0,1]$. More formally, they verify the following property: 

\begin{property}
\label{pr:DecreasingOptimalPathProperty}
Given a social network $G$ and a path $p=\{u_1,\dots,u_l\}$ in $G$, we have $\sigma^+(u_1,\dots,u_l)\geq \sigma^+(u_1,\dots,u_{l-1})$.
\end{property}


We then define $\sigma^+$  for any pair of user $(s,u)$ who are connected in the network by taking the maximal weight over all their connecting paths. More formally, we define $\sigma^+ (s,u)$ as
\begin{equation}
	\sigma^+ (s,u)= max_p \{\sigma^+ (p) ~|~  s\stackrel{p}{\leadsto}u\}.
	\label{equ:sigmaplus}
\end{equation}

Note that when the first candidate (multiplication) is used, we obtain the same aggregation scheme as in~\cite{Richardson03}, which is also  employed in~\cite{Schenkel08} in the context of top-$k$ network aware search.

\begin{example} 
In our running example, if we use multiplication in Eq.~(\ref{equ:sigmaplus}), for the seeker $Alice$, for $\alpha=0$ (hence exclusively social relevance), by Eq.(~\ref{equ:socfreq}) we  obtain the following values for social frequency: $SF_{Alice}(news)=\{D4:2.6, D2:1.01, D1:0.7, D6:0.6, D3:0.1, D5:0.05\}$ and $SF_{Alice}(site)=\{D4:1.11, D2:1.1, D3:0.9, D6:0.6, D1:0.05, D5:0.05\}$.
\label{ex:mulsflist} 
\end{example}

We argue next  that to all aggregation definitions that satisfy Property~\ref{pr:DecreasingOptimalPathProperty} and apply Eq.(\ref{equ:sigmaplus}) a greedy approach is applicable. This will allow us to browse the network of users on the fly, at query time,  visiting them in the order of their proximity with respect to the seeker. 

More precisely, by generalizing Dijkstra's algorithm~\cite{Dijkstra59}, we will maintain a max-priority queue, denoted $H$, whose top element $top(H)$ will be at any moment the \emph{most relevant unvisited user}\footnote{
Dijkstra's classic algorithm~\cite{Dijkstra59}  computes single-source shortest paths in a weighted graph without negative edges.}. A user is  \emph{visited} when her tagged items are taken into account for the top-$k$ result, as described in the following sections (this can occur at most once).
  At each step advancing in the network, the top of the queue is extracted (visited) and its unvisited neighbours (adjacent nodes) are added to the queue (if not already present) and are \emph{relaxed} . Let $\otimes$ denote the aggregation function over a path (one that satisfies Property~\ref{pr:DecreasingOptimalPathProperty}). Relaxation updates the best proximity score of these nodes, as described in Algorithm~\ref{alg:RelaxMax}.
\begin{algorithm} 
\caption{Relaxation}
\label{alg:RelaxMax}
\begin{algorithmic}	
	\IF{$\sigma^+(s,u) \otimes \sigma(u,v) > \sigma^+(s,v)$}
		\STATE $\sigma^+(s,v)= \sigma^+(s,u) \otimes \sigma(u,v)$
	\ENDIF
	\end{algorithmic}
\end{algorithm}

It can be shown by straightforward induction that this greedy approach allows us to visit the nodes of the network in decreasing order of their proximity with respect to the seeker, under any function for proximity aggregation that satisfies Property~\ref{pr:DecreasingOptimalPathProperty}.


We describe in the following section and in Section~\ref{sec:alg_general} how this greedy procedure for iterating over the network is used in our top-$k$ social retrieval algorithm. Without loss of generality, in the rest of the paper, consistent with social theories and with previous work on social top-$k$ search, proximity will be based on Candidate 1 (multiplication).

\section{Top-k Algorithm for $\boldsymbol{\alpha =0}$}\label{sec:algorithm0}
As the main focus of this paper is on the social aspects of search in tagging systems, we  detail first our top-$k$  algorithm, \textsc{TOPKS}, for  the special case when the parameter $\alpha$ is $0$. In this case, $fr(i~|~s,t)$ is simplified as 	
$$fr(i~|~s,t)=sf(i~|~s,t).$$
For each user $u$ and tag $t$, we assume a precomputed projection over the \emph{Tagged} relation for them, giving the items tagged by $u$ with $t$;  we call these the \emph{user lists}. No particular order is assumed for the items appearing in a user list.

We keep a list $D$ of top-$k$ candidate items, sorted in descending order by their minimal possible scores (to be defined shortly).  An item  becomes candidate when it is met for the first time in a $Tagged$ triple.

As usual, we assume that, for each tag $t$, we have an inverted list $IL(t)$ giving the items $i$ tagged by it, along with their term frequencies $tf(t, i)$\footnote{In \textsc{TOPKS}, even though the social frequency does not depend on  $tf$ scores, we will exploit the inverted lists and the $tf$ scores by which they are ordered, to better estimate score bounds. In particular, as detailed later, this allows us to achieve instance optimality.} in descending order of these frequencies. Starting from the topmost item, these lists will be consumed one item at  a time, whenever the current item becomes candidate for the top-$k$ result. By $CIL(t)$ we denote the items already consumed (as known candidates), by  $top\_item(t)$ we denote the item present at the current (unconsumed) position of $IL(t)$, and we use $top\_tf(t)$ as short notation for the term frequency associated with this item.

We  detail mostly the computation of  social frequency,  $sf(i~|~ u, t)$, as it is the key parameter in the scoring function of items. 
Since when $\alpha=0$ we do not use metrics that are tag-only dependent,  it is not necessary to treat each tag of the query as a distinct dimension and  to visit each in round-robin style (as done in the threshold algorithm  or in $\textsc{ContextMerge}$).  It suffices for our purposes to get at each step, for the currently visited user,  all the items that were tagged by her with  query terms (one user list for each term). 

 
For each tag $t_j\in Q$, by $unseen\_users(i,t_j)$ we denote the maximal number of yet unvisited users who may have tagged item $i$ with $t_j$. This  is initially set to the maximal possible term frequency of $t_j$ over all items (value that is available at the current position of the inverted list of $IL(t_j)$, as $top\_tf(t)$).  

Each time we visit a user $u$ who tagged  item $i$ with $t_j$ we can (a) update $sf(i ~|~ s, t_j)$ (initially set to $0$) by adding $\sigma^+(s,u)$ to it, and (b) decrement $unseen\_users(i,t_j)$. 

When $unseen\_users(i, t_i)$ reaches $0$,  the social frequency value $sf(i ~|~ s, t_j)$ is final. This also gives us a possible termination condition, as discussed in the following. 
  
  At any moment in the run of the algorithm,  the \emph{optimistic} score \textsc{MaxScore}$(i~|~s, Q)$ of an item $i$ that has already been seen in some user list  will be estimated using as social frequency for each tag $t_j$ of the query the following value:  
	$$top(H) \times unseen\_users(i,t_j) +  sf(i ~|~ s, t_j).$$
   Symmetrically,  the \emph{pessimistic} overall score, \textsc{MinScore}$(i~|~s,Q)$, is estimated  by the assumption that,  for each tag $t_j$, the current social frequency $sf(i ~|~ s, t_j)$ will be the final one.    
   The list of candidates $D$ is sorted in descending order by this lowest possible score.
  
  An upper-bound score on the yet unseen items,  \textsc{MaxScoreUnseen} is estimated using as social frequency for each tag $t_j$ the value $top(H) \times top\_tf(t))$.	
 \eat{ 
 The overall score estimations of a given item $i$, will then be:
  \begin{align}
  \nonumber \textsc{MaxScore}(i,Q)&=g(\textsc{MaxScore}(i~|~s,t_1),\dots,\textsc{MaxScore}(i~|~s,t_r))\\ 
  \nonumber \textsc{MinScore}(i,Q)&=g(\textsc{MinScore}(i~|~s,t_1),\dots,\textsc{MinScore}(i~|~s,t_r)) 
  \end{align}
 }  
 
 When the maximal optimistic score  of items  that are already in $D$ but not in its top-$k$ is less than the pessimistic score of the last element in the current top-$k$ of $D$ (i.e.,  $D[k]$),  
   the run of the algorithm can terminate, as we are guaranteed that the top-$k$ can no longer change. (Note however that at this point the top-$k$ items may have  only partial scores and,  if a ranked answer  is needed,  the process of visiting users  should continue.)
  
 	We present the flow of \textsc{TOPKS} in Algorithm~\ref{alg:soctopktrust}. Key differences with respect to  \textsc{ContextMerge}'s social branch are (i) the on-the-fly computation of proximity values, in lines 1-7 and 29-31 of the algorithm, and (ii) the consuming of inverted list positions,  when they become candidates, in lines 20-28.  For clarity, we first exemplify a  \textsc{TOPKS} run without the latter aspect (this would correspond to a \textsc{ContextMerge} run).  
 \begin{algorithm}\small
\caption{$\textsc{TOPKS}_{\alpha =0}$: top-$k$ algorithm for $\alpha=0$ }
	\begin{algorithmic}[1]\small
	\REQUIRE\emph{ seeker $s$, query $Q=(t_1, \dots, t_r)$}
	\FORALL{\emph{users $u$,  tags $t_j \in Q$, items $i$}}
		\STATE $\sigma^+(s,u)=-\infty$
		\STATE $sf(i ~|~ s, t_j)= 0$
		\STATE \emph{set $IL(t_j)$ position on first entry; $CIL(t_j)=\emptyset$}
	\ENDFOR
	\STATE $\sigma^+(s,s)=0$; $D=\emptyset$ (candidate items)  
	\STATE $H\gets$ \emph{max-priority queue of nodes $u$ (sorted by $\sigma^+(s,u)$),  initialized with  $\{s\}$}
	\WHILE{$H\neq \emptyset$}
	\STATE u=\textsc{extract\_max}(H);
		\FORALL{\emph{tags $t_j \in Q$, triples $Tagged(u,i,t_{j})$}}
			\STATE $sf(i ~|~ s, t_j)\gets sf(i ~|~ s, t_j) + \sigma^+(s,u)$
			\IF{$i \not \in D$} 
			\STATE \emph{add $i$ to $D$} 
			\FORALL{\emph{tags $t_l \in Q$}}
		\STATE $unseen\_users(i, t_l) \gets top\_tf(t_l)$(initialization)
	\ENDFOR
	 				\ENDIF	
					\STATE $unseen\_users(i, t_j) \gets unseen\_users(i,t_j) - 1$		
		\ENDFOR
		  \WHILE{\emph{$\exists t_j \in Q$ s.t. $i=top\_item(t_j) \in D$}}
	   \STATE {$tf(t_j,i) \gets top\_tf(t_j)$} \hspace{-0.7mm}($t_j$'s frequency in $i$ is now known)
	\STATE {\emph{advance $IL(t_j)$ one position}}
		\STATE {$\Delta \gets  tf(t_j,i) -  top\_tf(t_j)$ (the top\_tf drop)}
	\FORALL{\emph{items $i'\in D \setminus CIL(t_j)$}}
	\STATE $unseen\_users(i', t_j) \gets unseen\_users(i',t_j) - \Delta$
	\ENDFOR
	\STATE{\emph{add $i$ to $CIL(t_j)$}}
	   \ENDWHILE
		\FORALL{users $v$ s.t. $\sigma(u,v) \in E$}
			\STATE \textsc{relax}(u,v)
		\ENDFOR
		\IF{$\textsc{MinScore}(D[k],Q)> max_{l>k}(\textsc{MaxScore}(D[l],Q))$
		\textsc{and} \textsc{MinScore}$(D[k],Q)>\textsc{MaxScoreUnseen}$}
			\STATE \textbf{break}
		\ENDIF
	\ENDWHILE
	\RETURN $D[1], \dots, D[k]$ 
	\end{algorithmic}
\label{alg:soctopktrust}
\end{algorithm}    
    \begin{example}
Revisiting Example~\ref{ex:setting}, recall that we want to compute the top-$2$ items for the query $Q=\{news,site\}$ from $Alice$'s point of view. To simplify, let us assume that $score(i~|~u,t)=sf(i~|~u,t)$ and $g$ is addition. We consider next how the algorithm described above runs.
  
  At the first iteration of the line 8  loop in the algorithm, we visit $Bob$'s user lists, adding $D3$ to the candidate buffer. At the second iteration, we visit $Danny$'s user lists, adding $D2$ and $D4$ to the candidate buffer. At the third iteration ($Charlie$'s user list) we add $D6$ to the candidate list. $D1$ is added to the candidate list when the algorithm visits $Frank$'s user lists, at iteration 4. Recall that $top\_tf(news)=7$ and $top\_tf(site)=5$.
  
  The 6th iteration of the algorithm is the final one, visiting $George$'s user lists,  finding $D2$ tagged with $news,site$ and $D4$ tagged with $site$. $D4$ and $D2$ are the top-$2$ candidates, with $\textsc{MinScore}(D4,Q)$ $=2.61$ and $\textsc{MinScore}(D2,Q)=2.21$. The closest candidate is $D6$,  with  $\textsc{MinScore}(D6,Q)=1.2$ and $\textsc{MaxScore}(D6,Q)$ $=1.2+6\times0.1+4\times0.1=2.2$. Also,  $\textsc{MaxScoreUnseen}(Q)$ $=7\times0.1+5\times0.1=1.2$. Finally, $\textsc{MaxScore}(D6,Q)< \textsc{MinScore}(D2,Q)$ and since we have  $\textsc{MaxScoreUnseen}(Q)<\textsc{MinScore}(D2,Q)$, the algorithm  stops returning $D4$ and $D2$ as the top-$2$ items.
  \label{ex:topkssocinit}
  \end{example}
 We discuss next the interest of consuming of inverted list positions,  when these become candidates (illustrated in Example~\ref{ex:topksoclist}).  In lines 20-28,  we aim at keeping to a minimum the worst-case estimation of the number of unseen taggers.  More precisely, we test whether there are top-$k$ candidates $i$ (i.e., items already seen in user lists) for which the term frequency for some tag $t_j$ of $Q$, $tf(t_j, i)$,    is ``within reach'' as the one currently used (from $IL(t_j)$) as the basis for the optimistic  (maximal) estimate  of the number of yet unseen users who tagged  candidate items with $t_j$.  When such a pair $(i, t_j)$ is found, we can  do the following adjustments: 
  \begin{enumerate}
  \item refine the number of unseen users who tagged $i$ with $t_j$ from a (possibly loose) estimate to its \emph{exact} value; this is marked when $i$ is added to the $CIL$ list of $t_j$ (line 27), and from this point on the number of unseen users will only change when new users who tagged $i$ with $t_j$ are found (line 18).
  \item  advance (at the cost of a sequential access) beyond $i$ in the inverted list of $t_j$, to the next best item; this allows us to refine (at line 25) the estimates  $unseen\_users(i',t_j)$ for all candidates $i'$ for which the exact  number of users who tagged with $t_j$ is yet unknown.  
  \end{enumerate}
  \eat{
     Moreover, the items present in the user lists are the same as the items present in the inverted lists, and comparing a given item $i$ (encountered in the user lists) with the items present at the heads of the inverted lists has a constant cost. This allows us to add an optimization to the algorithm. We can compare the item $i$ with $head[t_j]$ for every $t_j\in Q$. If we find that $head[t_j]$ is the same item as $i$, then that means that $tf(t_j,i)=high(t_j)$, so we can update $pos\_tf(i,t_j)$. Finally, keeping the the head of $IL_{t_j}$ pointed to $i$ would mean that, for an item $d$ that doesn't know its $tf(d,t_j)$, $\textsc{MaxScore}(d~|~s,t_j)$ will be evaluated using the term-frequency of an already encountered item. Hence, incurring the cost of advancing $IL_{t_j}$ would arguably allow us to use a lower maximal term frequency, and thus refine the score estimations. }
    (We found in the experimental evaluation (Section~\ref{sec:results}) that this aspect has the potential to drastically improve the cost of the search.  Since  tf-values in inverted lists  fall quite rapidly in most practical settings, we witnessed significant cost savings, while using relatively few such list position increments.)
    \begin{example}
 Let us  now consider how the choice of advancing in the inverted lists when possible influences the number of needed iterations. At first,  $top\_tf(news)=7$, $top\_item(news)=D4$, and $top\_tf(site)=5$, $top\_item(site)=D2$.
    
    The first iteration only introduces $D3$ and thus we cannot advance in any of the two inverted lists. However, the discovery of $D2$ and $D4$ in step $2$ allows us to fix their exact tf values and advance the inverted lists. The new positions are: $top\_tf(news)=2$, $top\_item(news)=D1$, and $top\_tf(site)=1$, $top\_item(site)=D6$. $D6$'s discovery in iteration $3$ allows us to advance further in the inverted lists. 
    Finally, in step $4$, the discovery of $D1$ allows the algorithm to advance in the inverted lists to $top\_tf(news)=1$, $top\_item(news)=D5$, and $top\_tf(site)=1$, $top\_item(site)=D5$ (the only undiscovered item). This allows for some drastic score estimation refinements. We have the same top-$2$ candidates, $D4$ and $D2$ having $\textsc{MinScore}(D4,Q)=1.81$ and $\textsc{MinScore}(D2,Q)=1.21$. The closest item is again $D6$ having $\textsc{MinScore}(D6,Q)=\textsc{MaxScore}(D6,Q)=1.2$, since we know that we have visited all users who tagged $D6$. $\textsc{MaxScoreUnseen}(Q)=1\times0.3+1\times0.3=0.6$, since the maximal unseen document, $D6$ is tagged only once with each tag. $\textsc{MaxScore}(D6,Q)<\textsc{MinScore}(D2,Q)$ and $\textsc{MaxScoreUnseen}(Q)<\textsc{MinScore}(D2,Q)$ allows us to exit the loop,  two steps before the unrefined algorithm, returning the exact top-$2$: $D4$ and $D2$.

    \label{ex:topksoclist}
    \end{example}    


\noindent We can prove the following property of our algorithm: 

\begin{property}
\label{pr:DecreasingVisitProperty}
	 For a given seeker $s$,  $\textsc{TOPKS}_{\alpha =0}$ visits the network in decreasing order of the $\sigma^+$ values with respect to $s$.
\end{property}
As a corollary of Property~\ref{pr:DecreasingVisitProperty},  we have that $\textsc{TOPKS}_{\alpha =0}$ visit users who may be relevant for the query in the same order as \textsc{ContextMerge}~\cite{Schenkel08}. More importantly, we prove in  Section \ref{sec:optim}  that our algorithm  visits  as few users as possible, i.e., it is  instance optimal with respect to this aspect.  Moreover, the experiments show that  \textsc{TOPKS} can drastically reduce the number of visited user lists in practice (see Section \ref{sec:results}).
\subsection{Instance Optimality of $\boldsymbol{\textsc{TOPKS}_{\alpha=0}}$}
\label{sec:optim}
We will use the same definition of instance optimality as in~\cite{Fagin01}.  For a class of algorithms \textbf{A},  a class of legal inputs (instances)   \textbf{D},  $cost(\cal A, \cal D)$ denotes the cost of running algorithm $\cal A\in \textbf{A}$ on input $\cal D\in \textbf{D}$. An algorithm $\cal A$ is said to be \emph{instance optimal} for its class \textbf{A} over inputs \textbf{D} if for every $\cal B\in \textbf{A}$ and every $\cal D\in \textbf{D}$ we have $cost(\cal{A}, \cal{D})$=$O(cost(\cal{B}, \cal{D}))$.

Let $c_{UL}$ be the abstract cost of accessing the user list - a process which involves the relatively costly operations of finding the proximity value of the user and retrieving the items tagged by the user with query terms -  and let $users({\cal A, D})$ be the number of total user lists needed for establishing the top-$k$ for algorithm ${\cal A}$ on input ${\cal D}$. Let $c_{S}$ be the abstract cost of sequentially accessing the data in $IL_t$,  and let $seqitems({\cal A, D})$ be the total number of  sequential accesses to $IL$ for algorithm ${\cal A}$ on input ${\cal D}$. In practice, $c_{UL}\gg c_{S}$ is a reasonable assumption, hence, for two algorithms $\cal{A}$ and $\cal{B}$, we have

 $$\frac{users({\cal A,D })\times c_{UL} + seqitems({\cal A,D})\times c_{S}}{users({\cal B,D})\times c_{UL} + seqitems({\cal B,D})\times c_{S}}\approx \frac{users(\cal{A,D})}{users(\cal{B,D})}.$$

Therefore,  for a fair cost estimate in practical social search settings, a reasonable assumption is to consider 
$$cost({\cal A, D})=users({\cal A,D}).$$

Let us now define  the class of ``social'' algorithms \textbf{S} to which both $\textsc{TOPKS}_{\alpha=0}$ and \textsc{ContextMerge} (when $\alpha=0$) belong. These algorithms correctly return the top-$k$ items for a given query $Q$ and seeker $s$,  they do not use random accesses to $IL(t)$ indexes in order to fetch a certain $tf$ value, and they do not include in their working buffers (e.g., candidate buffer $D$) items that were not yet encountered in the user lists. The last assumption could be seen as a ``no wild guess'' policy, by which the algorithm cannot guess that an item might  be encountered in some later stages.    This is  a  reasonable assumption in practice, as the number of items needed for computing a top-$k$ result for a given seeker should in general be  much smaller than the total number of items tagged by query terms.

The class \textbf{D} of accepted inputs consists of the inputs that respect the setting described in Section~\ref{sec:setting}.


\begin{theorem}
	$\textsc{TOPKS}_{\alpha=0}$ is instance optimal over \textbf{S} and \textbf{D}, when the cost is defined as $cost({\cal A, D})=users({\cal A,D})$.
	\label{th:topks_soc_instopt}
\end{theorem}
The optimality proof is given in Appendix~\ref{sec:proof}.

\eat{
Due to space constraints, the proof can be found in the companion technical report~\cite{ManiuCautis11}.
}

\section{Algorithm for The General Case}\label{sec:alg_general}

For the general case, in which $\alpha \in [0,1]$, we adapt the \textsc{ContextMerge}~\cite{Schenkel08} algorithm to include the on-the-fly processing of user proximities. 

At each iteration, the algorithm can alternate, by calling \textsc{ChooseBranch()},   between two possible execution branches: the \emph{social branch} (lines 8-31 of Algorithm~\ref{alg:soctopktrust}) and the \emph{textual branch}, which is a direct adaptation of NRA.

As in the exclusively social setting of the previous section, we will read term frequency scores $tf(t_j,i)$ from the inverted lists, on a per-need basis, either as in line 21 of $\textsc{TOPKS}_{\alpha=0}$, or when advancing on the textual branch.  Initially, all unknown tf-scores are assumed to be set to $0$.  

The optimistic overall score  \textsc{MaxScore}$(i,Q)$ of an item $i$ that is already in the candidate list $D$ will now be computed by setting $fr(i ~|~ s, t)$, defined in Eq.~(\ref{equ:socfreq}), to 
\begin{eqnarray*}
fr(i ~|~ s, t) & =  & (1- \alpha) \times top(H) \times unseen\_users(i,t) +  (1-\alpha)\times sf(i ~|~ s, t) + \\
& & \alpha\times max(~tf(t,i), ~  top\_tf(t)~). 
 \end{eqnarray*}
 The last term accounts for the textual weight of the score, and uses either the exact term frequency (if known), or an upper-bound for it (the score in the current position of $IL(t)$).

Symmetrically, for the pessimistic overall score \textsc{MinScore}$(i,Q)$, the frequency  $fr(i~|~u,t)$  will be computed as  
\begin{eqnarray*}
fr(i ~|~ s, t) & =  &  (1-\alpha)\times sf(i ~|~ s, t) +  \alpha\times max(~tf(t,i) ~, ~  partial\_tf(t)~),
 \end{eqnarray*}
 where $partial\_tf$ represents the count of visited users who tagged $i$ with $t_j$, which is used as lower-bound for $tf(t_j,i)$ when this is not yet known.


The upper-bound for the score on the yet unseen items,  \textsc{MaxScoreUnseen}, is estimated using as overall frequency for each tag $t_j$ the following value:
$$fr(i ~|~ s, t) =\alpha\times top\_tf(t) + (1-\alpha)\times top(H) \times top\_tf(t)).$$

We present the flow of the general case algorithm in Algorithm~\ref{alg:gentopktrust}. Method \textsc{Initialize()} amounts to lines 1-6 of $\textsc{TOPKS}_{\alpha=0}$, and method \textsc{ProcessSocial()} amounts  to lines 8-31 of $\textsc{TOPKS}_{\alpha=0}$ (modulo the straightforward adjustment for the count $partial\_tf$). 

The difference between the $\alpha=0$ case and the general case is the processing of the inverted lists (textual branch), which is done as in the NRA algorithm (see lines 7-13 of Algorithm~\ref{alg:gentopktrust}). We discuss how the choice of the branch to be followed is done, by the  \textsc{ChooseBranch()} subroutine, in Section~\ref{sec:choosing}.

\subsection{Choosing between the social and textual branches}
\label{sec:choosing}
The  $\textsc{TOPKS}_{\alpha=0}$ algorithm,  in which only the social branch matters, is instance optimal (see Theorem~\ref{th:topks_soc_instopt}), with the cost being estimated as $users(\textsc{TOPKS}_{\alpha=0},{\cal D})$. As the NRA algorithm~\cite{Fagin01}, when only the textual branch matters,  $\textsc{TOPKS}_{\alpha=1}$ is instance optimal, with the cost being estimated as $seqitems(\textsc{TOPKS}_{\alpha=0}, {\cal D})$. 

When $\alpha$ is not one of the extreme values, under a cost function as a combination of the two above, of the form   
$$users(\textsc{TOPKS}_{\alpha=0}, {\cal D})\times c_{UL} + seqitems(\textsc{TOPKS}_{\alpha=1},{\cal D})\times c_{S},$$ 
a key role for efficiency is played by   \textsc{ChooseBranch()}.

In~\cite{Schenkel08}, the choice between the textual branch or the social one was done by estimating the maximum potential score of each, in round-robin manner over the query dimensions. For a query  tag $t_j$, the maximal contribution of the social branch would be estimated as
$\textsc{MaxSocial}(t_j)=(1-\alpha)\times max\_tf(t_j)\times top(H)$, 
where $max\_tf(t_j)$ is the maximum tf for  $t_j$ (i.e., the number of taggers for the item that has been tagged the most with $t_j$).  For the textual part, the maximal potential contribution would be estimated by setting
$\textsc{MaxTextual}(t_j)= \alpha \times top\_tf(t_j)$.  Then,  if $\textsc{MaxSocial}(t_j)>\textsc{MaxTextual}(t_j)$ the  social branch was chosen, otherwise the textual branch is chosen.

\begin{algorithm}
	\caption{\textsc{TOPKS}: top-$k$ algorithm for the general case}
	\begin{algorithmic}[1]\small
	\REQUIRE seeker $s$, query $Q=(t_1, \dots, t_r)$
	\STATE \textsc{Initialize()}
	\WHILE{$H\neq \emptyset$}
		\STATE \textsc{ChooseBranch()}
		\IF{\emph{social branch}}
			\STATE \textsc{ProcessSocial()}
		\ELSE
			\FORALL{\emph{tags $t_j\in Q$, item $i=top\_item(t_j)$}}
				\IF{$i\not\in D$}
					\STATE \emph{add $i$ to $D$ and $CIL(t_j)$}
				\ENDIF
				\STATE $tf(t_j,i)\gets top\_tf(t_j)$
				\STATE \emph{advance $IL(t_j)$ one position}
			\ENDFOR
		\ENDIF
		\IF{$\textsc{MinScore}(D[k],Q)>max_{l>k}(\textsc{MaxScore}(d[l],Q)$
		\textsc{and} \textsc{MinScore}$(D[k],Q)>\textsc{MaxScoreUnseen}$}
			\STATE \textbf{break}
		\ENDIF
	\ENDWHILE
	\RETURN $D[1], \dots, D[k]$
	\end{algorithmic}
	\label{alg:gentopktrust}
	\end{algorithm}

We use a different heuristics for the branch choice. At any  point in the run of  \textsc{TOPKS}, unless termination is reached,  we have at least one item $r$ with $\textsc{MaxScore}(r,Q)\geqslant\textsc{MinScore}(D[k],Q)$.  We consider the item  $r=D[argmax_{l>k}(\textsc{MaxScore}(D[l],Q)]$, which has the 
highest potential score, and we choose the branch that is the most likely to refine $r$'s score (put otherwise, the branch that counts the most in the \textsc{MaxScore} estimation for $r$). The intuition behind this branch choice mechanism is that it is more likely to advance the run of the algorithm closer to termination. 

For each tag $t_j \in Q$,  we set $\textsc{MaxTextual}(t_j)$ to $\alpha \times top_tf(t_j)$ if the term frequency $tf(t_j,r)$ is not yet known, or to $0$ otherwise. For the social part of the score,  we set 
$$\textsc{MaxSocial}(t_j)=(1-\alpha)\times unseen\_users(t_j,r)\times top(H).$$ 

Then, we follow the social branch if, for at least one of the tags, \textsc{MaxSocial} is greater than \textsc{MaxTextual}. 

Note that we deal with  the tags of the query ``in bulk'', and advance simultaneously on their inverted lists when the textual branch is followed. 

\textbf{Remark.} We have adopted so far a ``disjunctive'' interpretation for queries, in which items can score on each tag-dimension individually. However, our approach can be adapted in straightforward manner to a ``conjunctive'' interpretation: the pessimistic score should be maintained at $0$ until the item's scores -- at least partial ones -- are known for all tags.

\section{Efficiency by Approximation}
\label{sec:approx}
The algorithm described in the previous section is sound and complete, and requires  no prior (aggregated) knowledge on the proximity values with respect to a certain seeker (e.g., statistics); this was also the assumption in~\cite{Schenkel08}'s  \textsc{ContextMerge} algorithm. Moreover, it is instance optimal in the exclusively social setting (our main focus in this paper) with respect to the number of visited users.  While we  improve the running time in both this setting and the general one (more on experimental results in Section~\ref{sec:results}), in practice, however,  the search may still visit a significant part of the user network   and their item lists before being able to conclude that the top-k answer can no longer change. \eat{Note that, to test termination, we used  the tightest possible conditions (which might indeed be met in a social tagging configuration), when no  ``aggregated'' knowledge on how the proximity vector might look like is available (this was also the assumption in~\cite{Schenkel08}'s  \textsc{ContextMerge} algorithm).}

But if some statistics about proximity are known at query time (i.e., on how the values in a proximity vector variate from the most relevant user to the least relevant one), this may enable us to use more refined termination conditions, and thus to minimize the gap between the step at which the final top-k has been established and the actual termination of the algorithm. 
Indeed,  the  experiments we performed on Del.icio.us data 
showed that, in average, the last top-k change occurs much sooner, hence \emph{there is a clear opportunity to stop the browsing of the network earlier.}



\begin{figure}
	\centering
	\includegraphics[scale=0.6]{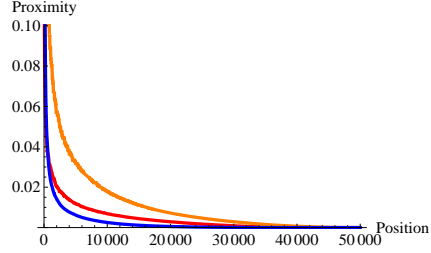}
	\caption{Examples on the evolution of proximity values.}
	\label{fig:proxvects}
\end{figure}

We take  a first step in this direction, discussing two possible approaches for using  score estimations based on proximity statistics,   which trade accuracy for efficiency (in terms of visited users). 
More specifically, in Algorithm~\ref{alg:gentopktrust}, the \textsc{MaxScore}, \textsc{MaxScoreUnseen} and \textsc{MinScore} bounds have all used the safest possible values for the  proximities of yet unseen users:  either the top (maximum) value of the max-priority queue ($top(H)$) for the first two bounds,  or its minimal possible value (zero) for the third one. In practice, however, any of these extreme configurations is rarely met.  For illustration, we give in Figure~\ref{fig:proxvects}  the  proximity vectors for some randomly sampled users. Observe that these  fall rapidly, and this may be  the case in many real-world similarity or proximity networks. 

Hence one possible direction for reducing the number of visited users is to pre-compute and materialize a high-level description (more or less complex, more or less accurate) of users' proximity vectors (of their distribution of values). 
This would allow us to use a tighter estimation for the remaining (unseen) users, instead of uniformly associating them the extreme score ($top(H)$ or $0$).  In doing so, we may obviously introduce approximations in the final result, and our approximate techniques provide a  trade-off between accuracy drop on one hand and  negligible memory consumption and reduced running time on the other hand. 

\subsection{Estimating bounds using mean and variance\label{sec:mvar}}
We first consider as a  proximity vector description one  that  is  very concise yet generally-applicable and effective,  keeping for a given seeker two parameters:  
the \emph{mean} value of the proximities in the vector and the \emph{variance} of these values. We adopt here the simplifying assumption  that the
values in the seeker's vector are independent, essentially interpreting the proximity vector as a random one. 

At any step in the run of the algorithm, using the mean and variance, for the remaining (yet unvisited) $unseen\_users(i,t)$  for a given item $i$ and tag $t \in Q$, we can 
derive (a) lower bounds for the average of their proximity values,  for \textsc{MinScore} estimations, or (b) upper bounds for the average of their proximity values, for \textsc{MaxScore} estimations.  The guarantees  of these bounds can be controlled  (in a probabilistic sense)  via a precision parameter  $\delta \in (0, 1]$, by which  lower values lead to higher precision and $1$ leads to a setting with no guarantees. 

More precisely, let $p$ be the current position in the proximity vector 
and let $\sigma^+_{p:}(s)$ be the vector containing the remaining (unseen) values of $\sigma^+(s)$. Knowing the overall mean and variance of the  entire proximity vector $\sigma^+(s)$, and having the proximity values seen so far (denoted  $\sigma^+_{0:p}(s)$), we can easily compute the average and variance of   the remaining proximity values   (those in $\sigma^+_{p:}(s)$).
\eat{
\vspace{-2mm}{\scriptsize $$Exp[\sigma_{p:}^+] = \frac{n\times Exp[\sigma^+] - p \times Exp[\sigma_{0:p}^+]}{n-p},$$
$$Exp[(\sigma_{p:}^+)^2]=\frac{n(Exp[\sigma^+]^2+Var[\sigma^+])-p(Exp[\sigma_{0:p}^+]^2+Var[\sigma_{0:p}^+])}{n-p},$$
$$\tiny{Var[\sigma_{p:}^+]=Exp[(\sigma_{p:}^+)^2]-Exp[\sigma_{p:}^+]^2}.$$}
}


Then, the mean and variance of the average of $unseen\_users(i,t)$ randomly chosen proximity values from the remaining ones can be obtained as follows:\footnote{This is possible under independence assumptions that may not entirely hold, but  turn out to be reasonable in practice (see Section~\ref{sec:results}).}  
\begin{eqnarray*}
 \vspace{-2mm}
Exp[\sigma^+_{p:},unseen\_users(i,t)] & = & E[\sigma^+_{p:}],\\
Var[\sigma^+_{p:},unseen\_users(i,t)] & = & \frac{Var[\sigma^+_{p:}]}{unseen\_users(i,t)}.
\end{eqnarray*}
\eat{where $E[\sigma^+_{p:}]$ and $Var[\sigma^+_{p:}]$ represent the mean and variance of $\sigma^+_{p:}$, respectively.}
When the input query  contains more than one tag, its size  $|Q|$ needs to be taken into account in the estimations. In order to avoid computational overhead, we uniformly chose a non-optimal per-tag probabilistic parameter $\delta'$  that ignores per-tag score distributions, as follows:
\begin{equation}
  \delta'=1-(1-\delta)^{1/|Q|}.
  \label{equ:qbound}
\end{equation}
$\textsc{EstMax}(p,\delta)$ represents, for each query tag,  the upper bound of the expected value of the average of $unseen\_users(i,t)$ values drawn from $\sigma^+_{p:}(s)$, which holds with probability at least $1-\delta'$.  Similarly, $\textsc{EstMin}(p,\delta)$ represents the lower bound of the expected value of the average of $unseen\_users(i,t)$ values drawn from $\sigma^+_{p:}(s)$, which holds  with probability at least $1-\delta'$.  For estimating \textsc{MinScore} when $i\not\in CIL(t)$, the fact that we have no information about  the difference between $tf(i,t)$ and $partial\_tf(t,i)$ (the users who  tagged item $i$ with $t$ so far) means that
we cannot assume that  other users may have tagged $i$, so we keep this estimation as in the initial (exact) algorithm.

By using Chebyshev's inequality, these bounds can be computed as follows:
\begin{eqnarray*}
\textsc{EstMax}(p,\delta) & = & E[\sigma^+_{p:}(s)]+\sqrt{\frac{Var[\sigma^+_{p:}(s)]}{unseen\_users(i,t)\times \delta'}}\\
\textsc{EstMin}(p,\delta)& = & E[\sigma^+_{p:}(s)]-\sqrt{\frac{Var[\sigma^+_{p:}(s)]}{unseen\_users(i,t)\times \delta'}}
\end{eqnarray*}
We give the score estimations, changed by generalizing the proximity estimations, in Table~\ref{tab:scorest}.  We present in the experimental results the effect of this approximate approach on running time,  showing significant overall improvement.   
In our experiments, even for $\delta=0.9$, the  returned top-$k$ answers had reasonable precision levels (around $90\%$).
 
We discuss in the next section another approach for tighter score estimates, using more detailed descriptions of proximity vectors. We conclude this section with a discussion on how these concise descriptions of proximity vectors could be maintained up-to-date in dynamic environments, in Section~\ref{sec:maintain-approx}.

\begin{table*}\scriptsize
    \centering
    \caption{Optimistic and pessimistic estimations of $fr(i~|~t,u)$ at step $p$ (general case).\label{tab:scorest}}
    \begin{tabular}{ccc}
		\toprule
		\textbf{score}		& $\boldsymbol{i \in CIL(t)}$ & \textbf{estimation} \\
		\midrule
		\textsc{MinScore}(i,t) 	& yes	& $\alpha\times tf(i,t) + (1-\alpha)\times (sf(i ~|~ s, t)+\boldsymbol{\textsc{EstMin}(p,\delta)} \times unseen\_users(i,t))$\\
						& no		& $\alpha\times partial\_tf(t,i)+(1-\alpha)\times sf(i ~|~ s, t)$\\
		\midrule
		\textsc{MaxScore}(i,t)  	& yes	&$\alpha\times tf(i,t) + (1-\alpha)\times (sf(i ~|~ s, t) +\boldsymbol{\textsc{EstMax}(p,\delta)} \times unseen\_users(i,t))$\\
						& no		& $\alpha\times top\_tf(t)+ (1-\alpha)\times (sf(i ~|~ s, t) + \boldsymbol{\textsc{EstMax}(p,\delta)} \times unseen\_users(i,t))$\\
		\midrule
		\textsc{MaxScoreUnseen}(t)	& 		& $\alpha\times top\_tf(t) + (1-\alpha)\times  \boldsymbol{\textsc{EstMax}(p,\delta)} \times top\_tf(t)$\\
		\bottomrule
    \end{tabular}
    \end{table*}

\subsection{Estimating bounds using histograms}\label{sec:hist}

The advantage of the approach described the previous section is twofold:  low memory requirements and estimation bounds that are applicable for  any value distribution. However,  it may offer estimation bounds that are  too loose in practice, and hence not reach the full potential for efficiency of approximate score bounds.  To address this issue, we can imagine -- as a compromise between keeping only these two statistics and keeping the entire pre-computed proximity vector --  an approach in which we describe the distribution at a finer granularity, based on \emph{histograms}. 

More precisely, for a seeker $s$,  we denote this histogram as $h(\sigma^+(s))$. It consists of $b$ buckets, each bucket $b_i$, for $i\in\{1,\dots,b\}$, containing $n_i$ items in the interval $(low_i,high_i]$ (the  $0$ values are assigned to bucket $b$). 
 Then, the probability that there exists a proximity value $x$ greater than $low_i$, knowing the histogram $h(\sigma^+(s))$,  is  
$$Pr[x>low_i~|~h(\sigma^+(s))]=\sum_{j=1}^{i}n_j/n.$$

At any step $p$ in the run of the algorithm, we maintain a partial histogram denoted as $h(\sigma^+_{p:}(s))$, obtained by removing from $h(\sigma^+(s))$  the $p$ already encountered proximity values.

Similar to the previous approach, we can drill down the overall $\delta$ parameter to a $\delta'$ one for each query tag. Then, $\textsc{EstMax}(p,\delta)$ can be given by the minimal value in the partial histogram, such that the resulting estimation of $\textsc{MaxScore}(i,t)$ holds with at least probability $1-\delta'$. Conversely, $\textsc{EstMin}(p,\delta)$ is given by the maximal value in the partial histogram, such that the resulting estimation of $\textsc{MinScore}(i,t)$ holds with at least probability $1-\delta'$.

In manner similar to Eq.(\ref{equ:qbound}), we need to take into account the fact that a number of $unseen\_users(i,t)$ such estimated values lead to an overall approximate estimation, for both $\textsc{EstMin}$ and \textsc{EstMax}. Therefore,  each of these values is uniformly estimated using a stronger probabilistic parameter $\delta''(i,t)$, depending on $unseen\_users(i,t)$, as follows:
$$\delta''(i,t)=1-(1-\delta'')^{1/unseen\_users(i,t)}.$$

Formally, having $h(\sigma_{p:}^+(s))$ and $\delta''(i,t)$, we estimate $\textsc{EstMax}(p,\delta)$ and $\textsc{EstMin}(p,\delta)$ as follows:
$$\textsc{EstMax}(p,\delta) = min \{low_i~|~Pr[x>low_i~|~h(\sigma^+_{p:}(s))]\leq\delta''(i,t)\},$$
$$\textsc{EstMin}(p,\delta) = max \{low_i~|~Pr[x>low_i~|~h(\sigma^+_{p:}(s))]\geq1-\delta''(i,t)\}.$$


The space needed for keeping such  histograms is linear in the number of users and buckets.  For instance, by setting the latter using the square-root choice, the memory needed is $O(n^\frac{3}{2})$. Also, as a consequence of the on-the-fly computation of  proximity values, we can easily update the histogram of the seeker by merging the partial, ``fresh''  histogram obtained in the current run (until  termination)  with the remaining values from the existing (pre-computed) histogram.

\subsection{Maintaining the description of the proximity vector}
\label{sec:maintain-approx}
Since social tagging applications are highly dynamic in nature, we need to take into account the fact that the statistics we keep are  likely to change quite often.  While we can hope that mean, variance and even histogram descriptions are less subject to change than individual proximity values, we should still strive to maintain these statistics as fresh as possible.   Recomputing them from scratch, at certain intervals, is an obvious option to consider, though one that may still be too  expensive, knowing that we want to avoid keeping the $n\times n$ materialized proximity matrix, as well as na\"ive re-computation of mean and variance pairs.  

A more suitable alternative would be to rely on approximate techniques for maintaining  a fully dynamic all-pairs shortest path information (APSP) in the network.  
Since our  proximity metric relies on  path multiplication,  we can reformulate the computation of proximity values into a problem of computing shortest paths in  a  network with (a) the same  set of vertices and edges, and (b) edge weights valued $w(u,v)=-\log \sigma(u,v)$, where $\sigma(u,v)$ is the user proximity  from the original network. 


A ($2+\epsilon$)-approximate algorithm was given in~\cite{Bernstein09}, which handles fully-dynamic updates in a graph in $\tilde{O}(e)$ (almost linear)  time. It exhibits a query time of $O(\log \log \log n)$ (the query returns an estimation of the shortest distance between two nodes), without the need of keeping a distance matrix. We could directly  rely on  this algorithm in the transformed $-\log\sigma(u,v)$ graph.  Mapping back the distances thus queried to our setting would give us an estimation $\sigma^+_{est}(u,v)$ that verifies the inequality:
$$\sigma^+(u,v)\geqslant\sigma^+_{est}(u,v)\geqslant\sigma^+(u,v)^{2+\epsilon}.$$

For a given seeker $s$, we could thus compute an approximation of its proximity vector in $O(n\log \log \log n)$ time, and then compute the approximate statistics  efficiently. 


\section{Scaling and performance}
\label{sec:performance}
We argue in this section that, in a real-world setting, our algorithm \textsc{TOPKS} outperforms the one from existing literature both in terms of memory requirements and execution time. We discuss its practical impact in experiments in Section~\ref{sec:results}.

Let us consider, as an illustrating example, one of the most popular bookmarking applications, Del.icio.us, which currently has probably around $10^7$ users.  \eat{ (Twice less  than that was reported in $2008$ for Del.icio.us; the image tagging system Flickr has a user base of comparable size.) }Unsurprisingly, this social network  is quite sparse, with an average degree of about $100$. 
If a similar graph configuration would be maintained when weights (the $\sigma$ function) are associated to the edges of the network (e.g., based on tagging proximity or some other measure) the size of an index that would precompute the extended proximity value for each pair of connected users in the network (the $\sigma^+$ function) would be roughly of $700$ terabytes  (i.e., $(10^7)^2 \times 7$ bytes, considering that $3$ bytes are necessary for an user Id and  $4$ bytes are necessary for the float value of proximity).  On the other hand, the weighted graph would require memory space of roughly $7$ gigabytes (as $10^7 \times 100 \times 7$ bytes), and could easily fit in the RAM space of an average commodity workstation.\footnote{We stress that, for the sake of generality, this is not assumed nor exploited in our algorithms, and is not accounted for in the experimental results for \textsc{TOPKS} (in both abstract cost and running time).} More, existing techniques for network compression~\cite{Chierichetti09}  might allow us to reduce the space required to store the network by a factor of $10-15$ while still supporting efficient updates and random access on compressed data. 

The difference in memory requirements for the two alternatives becomes much more drastic when assuming a user base of the order of Facebook's social network, which currently consists of roughly  $7\times 10^8$ users (and is still growing at a fast pace). Precomputed lists for extended proximity go up to about $400$ petabytes of memory space, while the network itself requires only about  $400$ gigabytes.  The space needed to store the network can further  decrease to fit RAM capacity that moderate commodity servers can provide today,    if considering the compression techniques mentioned previously. 

We next  discuss general performance aspects, which in practice may be as impacting as the memory and updatability advantages that our algorithm presents.

 Let $n$ denote the number of users and let $e$ denote the number of edges in the network.  We assume without loss of generality that  the query consists of a single tag  (for  multiple-tag queries, all dimensions can share the results of a single $\sigma^+$ computation).

 \begin{table}\small
     \centering
     \caption{\label{tab:accanalysis}Computational costs for processing a query $Q$, when $\alpha=0$.}
	\begin{tabular}{cccc}	
	\toprule	
		
	\textbf{Algorithm}	& \multicolumn{2}{c}{\textbf{Disk access}} &  \textbf{RAM access}\\
	\cmidrule(rl){2-3}
		& \textbf{RA} & \textbf{SA} & \\
	\midrule
	 \textsc{ContextMerge} & 1 & $n$ & $(|Q|-1)\times n$\\
	$\textsc{TOPKS}_{\alpha=0}$ & 0 & 0 & $O(n\lg n + e)+(|Q|-1)\times n+n+e$ \\
	\bottomrule
	\end{tabular}
\end{table}

For our algorithm,  let us  assume that the social network resides in main memory, e.g., by means of adjacency lists: for each vertex, we have a list of its neighbors and their associated weights (we can safely assume the list comes presorted descending  by weight).  For one top-$k$ query execution, we will need at most  $n+e$ operations to visit the entire network  (we are guaranteed to take each vertex only once). 
 For the proximity computation we can use a Fibonacci-heap based max-priority queue, since our graph is likely to be very sparse~\cite{Mislove07}.  Each insertion into the heap takes $O(1)$ amortized time,  each extraction takes $O(\lg n)$ and each increase of a key (a relaxation step) takes $O(\lg n)$, for an overall queue complexity of  $O(n\lg n + e)$. 

  \textsc{ContextMerge} requires no computations for proximity at que\-ry time. However,  it uses  disk accesses to read the precomputed proximity values: one random access to locate the seeker's list  and $n$ sequential disk accesses to read this list. (It suffices to do this just for one query term,  and then keep and access a shared copy of this list in main memory.)


If we value the latency of a memory access as $1$ and the one of  a sequential disk access as $t$ (usually about five orders of magnitude slower than RAM access),  with minor simplifications,  our algorithm has the potential to perform better than \textsc{ContextMerge} when the following holds:
$t > \lg n +  \frac{e}{n}$. 
So the network sparseness  should verify the following inequality:
$$e <  n \times (t - \lg n),$$  
which is a very plausible assumption  in real applications.

A summary of this comparison on execution time is given in Table~\ref{tab:accanalysis}. 
Note that in this analysis we omitted initialization costs: the  overhead necessary for  \textsc{ContextMerge} to compute  $\sigma^+$ values for all  user pairs and the overhead to load in main-memory the social network,  for our algorithm. 


\section{Experimental Results}\label{sec:results}
 \noindent \textbf{Dataset and testing methodology.}  We have performed our experiments on a publicly available Del.icio.us dataset~\cite{Wetzker08}, containing $80000$ users tagging $595811$ items with $198080$ tags. As this  dataset does not  give information regarding links between users, we have generated three similarity networks:

 \begin{itemize}
 \item \emph{Item similarity network.} This network was constructed by computing the Dice coefficient of the common items bookmarked by any two users, resulting in a network of $49038$ users and $3329540$ links.
 \item \emph{Tag similarity network.} This network was generated by computing the Dice coefficient of the common tags used by any two users. Since this computation results in a  network that is too dense, we have filtered out the users who used less than $10$ distinct tags in their tagging activity. The final networks thus contains $40319$ users and $8335544$ links.
\item \emph{Item-tag similarity network.} This network was constructed by computing the Dice coefficient of the common items and tags bookmarked by any two users, resulting in a network containing $40353$ users and $1849898$ links.
 \end{itemize}

We computed the top-$10$ and top-$20$ answers, generating a number of $20$ two and three-tag semantically coherent queries, from tags that have a medium frequency (i.e., between $3000$ and $5000$ in our dataset). For each similarity network, $10$ random users were also randomly chosen in the role of  the seeker.
 
 Testing was performed using two ranking functions (the $h$-function from our model). The first one is the standard tf-idf ranking function:
 $$score(i~|~u,t)=fr(i~|~u,t)\times idf(t).$$
The second one is the BM15 ranking function used in~\cite{Schenkel08}: 
 $$score(i~|~ u, t)=\frac{(k1+1)fr(i~|~ u, t)}{k1+fr(i~|~ u, t)}\times idf(t),$$
where  inverse frequency $idf(t)$ is defined in standard manner as
	$$idf(t)=\log \frac{|{\cal I}| - |\{i ~|~Tagged(v,i,t)\}|+0.5}{|\{i ~|~ Tagged(v,i,t)\}|+0.5}.$$
and the aggregation function $g$ is summation. 

While these are two of the most commonly used ranking functions in IR literature, they have different properties  when used in approximate approaches as the ones we describe. More precisely, since tf-idf is  a linear function, both the maximal and minimal estimates over $fr$ scores lead to valid estimates for the overall scores. This is not necessarily the case for BM15: since it is a concave function, only the maximal overall score can be estimated. This was taken into account in the experiments.

We used a Java implementation of our algorithms,   on a machine with a 2.8GHz Intel Core i7 CPU, 8GB of RAM, running Ubuntu Linux 10.04 and  PostgreSQL 9.0.

As our focus  is on optimizing the social branch of the top-$k$ retrieval, we report here our results for $\alpha\in\{0,0.1,0.2,0.3\}$. As~\cite{Schenkel08}, multiplication over the paths was chosen as the proximity aggregation function, as the best suited candidate  for predicting  implicit similarities.

\textbf{Remark.} The relevance of personalized query results  is a topic that has been extensively treated (\cite{Wang10,Dou07,Konstas09}). It is not our focus here, and we interpret the relevance of  results as a consequence of the scoring functions $g$ and $h$. Moreover, the query itself could be viewed as  the result of a transformation using techniques such as query expansion.  The relevance of social search results was also extensively evaluated in~\cite{Schenkel08}, over Del.icio.us data, in a setting (including ranking model) similar to ours. We report, however, on two ground-truth experiments for evaluating the relevance of top-$k$ results for $\alpha=0$, at the end of this section.


\paragraph*{Efficiency results.} For the testing environment described previously, we report on efficiency  for both exact and approximate algorithms,  and on  precision for the latter. 
\eat{For efficiency, we focus on abstract cost, which is the standard approach for early-termination algorithms that depend on database accesses. For comparison, we also give in the technical report~\cite{ManiuCautis11} the corresponding running times, and one can notice there that  abstract cost  closely captures the actual performance of the algorithms.}

For efficiency, we report on two measures: the abstract cost of the algorithms and their wall clock running times. Abstract cost, which is the standard measure for early-termination algorithms that depend on database accesses, is computed as defined in Section~\ref{sec:choosing}, by choosing $c_{UL}$, the cost of accessing a user lists, as valued  $100$ (a very conservative upper-bound), and $c_{S}$, the cost of sequentially accessing an item in an inverted lists,  valued $1$. More formally, 
$$cost(A,D)=100\times users(A,D)+seqitems(A,D).$$ 
We  ignore differences in favor of \textsc{TOPKS} that are hard to account for, namely we do not distinguish between the user accesses by \textsc{ContextMerge} (which in a real setting would be to external memory) and the ones by  \textsc{TOPKS} (which would be to main memory).

\begin{figure*}
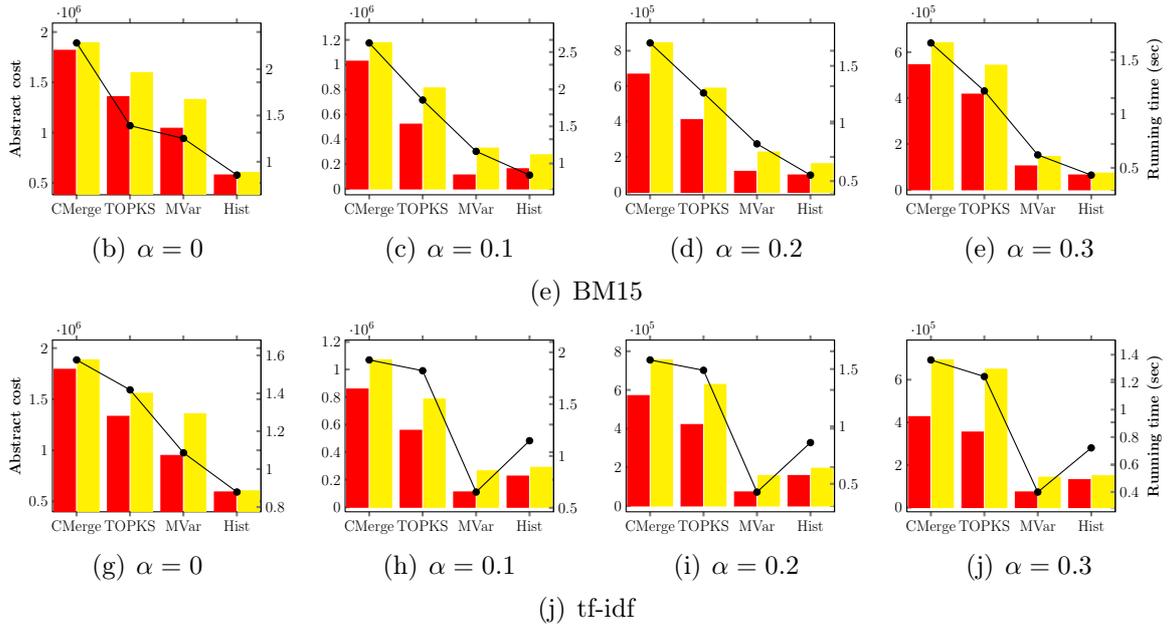
\small

\centering
\subfigure[BM15]{
\subfigure[$\alpha=0$]{

}
}
\caption{Abstract cost and running time comparison over the  tag-similarity network and the $f_{mul}$ proximity function (red: top-10, yellow: top-20).\label{fig:tagsim}}
\end{figure*}

Figures~\ref{fig:tagsim},\ref{fig:docsim} and \ref{fig:doctagsim} present the comparison of abstract costs and running times for the BM15 and tf-idf ranking functions, for each of the three similarity networks. In each subfigure, the first pair of columns gives the abstract cost of~\cite{Schenkel08}'s \textsc{ContextMerge} algorithm, the second pair of columns the one of \textsc{TOPKS}, the third pair of columns the cost of $\textsc{TOPKS}/MVar$ (approximate approach based on mean and variance of proximities, described in Section~\ref{sec:mvar}) and the fourth pair of columns the cost of $\textsc{TOPKS}/Hist$ (approximate approach based on  histograms, described in Section~\ref{sec:hist}). For each algorithm,  the average running times were recorded, and are represented by the black line in the plots (one dot indicates  the average running time between the top-10 and the top-20). 
One can notice there that abstract cost closely captures the actual performance of the algorithms. However, running time optimization was not the focus of the present work, and many alternatives remain to be explored in that direction (e.g., tuning the database).\footnote{Note that we cannot compare with~\cite{Amer-Yahia08}'s approach, as it only extends classic top-k retrieval by interpreting  user proximity  as a binary function (0-1 proximity), by which only users who are directly connected to the seeker can influence the top-k result.}

First, we can see that in general $\textsc{TOPKS}$ drastically  improves efficiency when compared to \textsc{ContextMerge}, in terms of  both running time and abstract cost. For example, in the item-tag similarity network, when $\alpha=0$, the running time and abstract cost are around 50\% of that of \textsc{ContextMerge}.
 
Moreover, our  approximate approaches lead to further improvements, which support the intuition that even  limited statistics (such as mean and variance) can render  the termination conditions more tight.

\begin{figure}
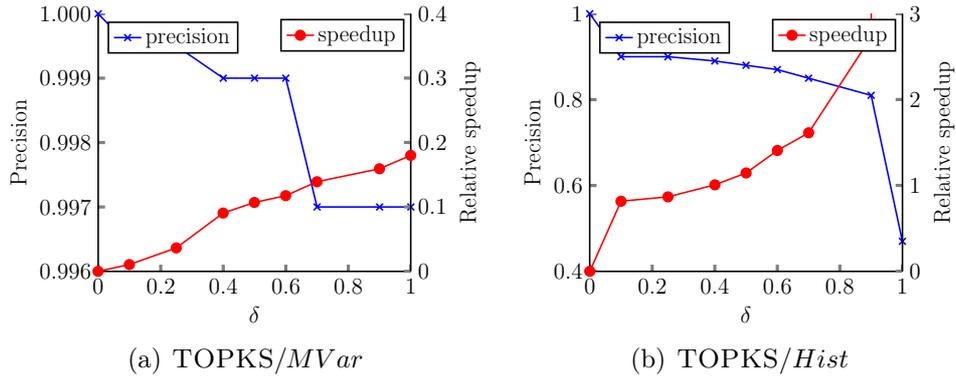

  \centering
  \subfigure[$\textsc{TOPKS}/MVar$]{
   %
    }
\caption{Precision rates versus speedup relative to \textsc{TOPKS}, when $\alpha=0$.\label{fig:speed}}
\end{figure}

\begin{figure}
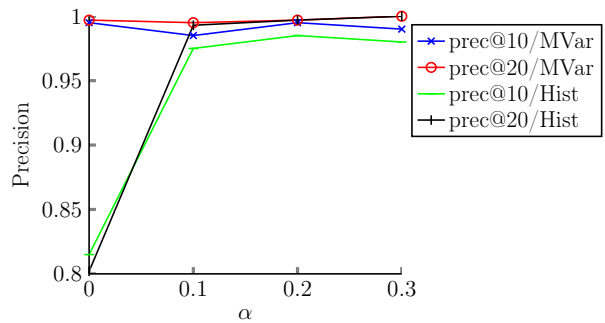

  \centering
   %
   \caption{Precision rates vs. $\alpha$, when $\delta=0.9$.\label{fig:prec}}
 \end{figure}

The abstract costs of $\textsc{TOPKS}/MVar$ and $\textsc{TOPKS}/Hist$ in the figure were obtained for the probabilistic threshold $\delta=0.9$. Even though this  represents a quite weak guarantee,  
we found that it  still yields  a  good precision/efficiency trade-off.   For a better understanding of this trade-off, we show  in Figure~\ref{fig:speed} the impact of $\delta$ on precision.
When $\alpha>0$, visiting the per-term inverted lists in parallel to the proximity vector helps in deriving tighter score bounds for unseen items, leading to a faster termination of the approximate approaches. These tighter score bounds also help in achieving better precision levels when $\alpha>0$, as Figure~\ref{fig:prec} shows.

Furthermore, our branch choice heuristic in $\textsc{TOPKS}$ (in both the exact and approximate variants)  brings significant improvements overall (for instance, consider the difference between the cost savings for $\alpha=0$ and $\alpha=0.1$, in the tag similarity network). 
Finding even more effective heuristics for this aspect of the algorithm remains  an interesting direction for future research. 

We discuss next how the instance optimality of $\textsc{TOPKS}_{\alpha=0}$ reflects in the performance results. Table~\ref{tab:optim} reports the number of visited users by  \textsc{ContextMerge} and $\textsc{TOPKS}_{\alpha=0}$ (columns \emph{users}), for the three similarity networks. One can see that $\textsc{TOPKS}_{\alpha=0}$ achieves good savings (in terms of  visited users), while relying only on very few sequential accesses in the inverted lists (column \emph{seqitems}).

\begin{table}\small
    \centering
\caption{\label{tab:optim}Comparison between \textsc{ContextMerge} and $\textsc{TOPKS}_{\alpha=0}$.}
\begin{tabular}{crrrr}
	\toprule
	\textbf{Network} & \multicolumn{2}{c}{\textsc{ContextMerge}} & \multicolumn{2}{c}{$\textsc{TOPKS}_{\alpha=0}$}\\
	\cmidrule(rl){2-3} \cmidrule(rl){4-5}
	& users & seqitems & users & seqitems \\
	\midrule
	item & 21878 & 0 & 15588 & 65 \\
	item-tag & 13028 & 0 & 6898 & 54 \\
	tag & 18718 & 0 & 15581 & 68 \\
	\bottomrule
\end{tabular}
\end{table}

Finally, we consider the impact of the probabilistic parameter $\delta$ on  precision and speedup in the approximate algorithms.  We define precision as the ratio between the size of the exact result (by  \textsc{TOPKS}) and the number of common items returned by the respective approximate approach and  \textsc{TOPKS}, i.e., 
$$precision=\frac{|T_{\textsc{TOPKS}/app}\cap T_{\textsc{TOPKS}}|}{|T_{\textsc{TOPKS}}|},$$
where $T_{\textsc{TOPKS}/app}$ is the set of items returned as top-k by the approximate algorithms (either $\textsc{TOPKS}/MVar$ or $\textsc{TOPKS}/Hist$), and $T_{\textsc{TOPKS}}$ is the set of items returned by the exact algorithm. 

The relative speedup is defined as $$speedup=\frac{cost(\textsc{TOPKS},D)}{cost(\textsc{TOPKS}/app,D)}-1.$$
We present in Figure~\ref{fig:speed} the results for both approximate approaches, $\textsc{TOPKS}/MVar$ and $\textsc{TOPKS}/Hist$. For $\textsc{TOPKS}/MVar$, one can notice  that $\delta$ has a limited influence on  precision  (with a minimum of $0.997$ for $\delta=1$), while ensuring reasonable speedup. The speedup potential is greater when using $\textsc{TOPKS}/Hist$ and histograms, while  reasonable precision levels are obtained (for instance,  precision of around $0.805$ when $\delta=0.9$, for a speedup of around $2.5$). For values of $\delta>0.9$, we notice however a rapid drop in precision.
The fact that $MVar$ achieves better precision than $Hist$ may seem counter-intuitive, since histograms give a more detailed description of proximity vectors. This difference in precision is due to looser bounds for $MVar$, as they directly influence the termination condition of the algorithm, result in a longer run and hence to better chances of returning a more refined top-$k$ results.

We also considered  the influence of the $\alpha$ parameter on  precision,  while setting the probabilistic parameter to $\delta=0.9$ (see Figure~\ref{fig:prec}). We have measured both $precision@10$ (i.e., when requesting the top-10) and $precision@20$ for both $\textsc{TOPKS}/MVar$ and $\textsc{TOPKS}/Hist$. We observed that the precision levels for $\textsc{TOPKS}/MVar$ are quite stable for all values of $\alpha$. For $\textsc{TOPKS}/Hist$, the lowest values of precision are witnessed when $\alpha=0$, but they stabilize to high values (above $0.97$) for $\alpha>0$.

\eat{Moreover,  even when ignoring the variance in our estimations, the precision remains high (line \emph{w/o var}).} 



\begin{figure*}\small
\centering
\subfigure[BM15]{
\subfigure[$\alpha=0$]{
%
}
}
\caption{Abstract cost and running time comparison over the  item-similarity network and the $f_{mul}$ proximity function (red: top-10, yellow: top-20).\label{fig:docsim}}
\end{figure*}

\begin{figure*}
\centering
\subfigure[BM15]{
\subfigure[$\alpha=0$]{
%
}
}
\caption{Abstract cost and running time comparison over the  item-tag similarity network and the $f_{mul}$ proximity function (red: top-10, yellow: top-20).\label{fig:doctagsim}}
\end{figure*}

\eat{
 \paragraph*{Scaling} We next look at how our approaches scale when one varies important parameters. To test this, we have varied two parameters, for $\alpha=0$: the number of desired results $k$, and the size of the network in number of users.

 We first show the scaling in terms of $k$.

\begin{figure*}
\centering
\subfigure{
%
}
{\tiny \ref{legsc}}
\caption{Abstract cost of algorithms when varying $k$.\label{fig:scaling_k}}
\end{figure*}
 }

\paragraph*{Evaluating relevance} We report now on two ``ground-truth'' experiments we have performed to test the bookmark prediction power of the exclusively social queries.

For the first experiment, we have selected $(user, tag)$ pairs from users that have bookmarked between 5 and 10 items using tags that were used globally at least 1000 times. The objective of this experiment was to estimate the power of personalized results to predict items that are tagged using relatively popular tags.

Then, 1000 pairs were randomly selected. For each pair and for $k\in\{1,2,5,7,10\}$, we computed the following top-$k$ result, using as query the tags corresponding to the distinct user-ids: the network-unaware top-$k$, and, setting the user-id as the seeker, the personalized top-$k$ (for $\alpha=0$) for each of the following aggregation functions: $f_{mul}$, $f_{min}$, and $f_{pow}$ with $\lambda\in\{1.1,2\}$. For each personalized query, the items belonging to the seeker were ignored (so as not to influence positively the precision of the results).

An item was considered as ``predicted'' if it appeared in the resulting top-$k$ and was also tagged by the seeker userid with the query tags. We traced the proportion of pairs for which at least one such item has been predicted.

The results are presented in Figure~\ref{fig:relev_exp1}. One can note that, for the item and item-tag similarity networks, personalization is considerably better at predicting bookmarked items than the ``global'' top-$k$, for all functions, except $f_{min}$ and, to a lesser extent, $f_{pow}$ ($\lambda=1.1$). Moreover, the tag similarity network seems to not be such a good predictor, no matter the personalization function used, as the other two networks. This might indicate the fact that, in the case of tag similarity, one needs to go beyond simple set similarities and include more complex relationships between tags, like synonymy and polysemy. 

For the second experiment, we have selected $(user, item, tag)$ triples resulting from items that have been tagged only by few people in the network (between 5 and 10). The objective of this experiment was to estimate the power of personalizing results to predict items that are unpopular, i.e., the ``long tail''. The tests and the measures tracked are identical to the setup of the first experiment. 

\begin{figure*}
\centering
\subfigure{
\begin{tikzpicture}[scale=0.55]
  \begin{axis}[
    title = {tag similarity},
    legend columns = -1,
    ylabel={\textbf{precision}},
    xlabel={\textbf{k}},
    ymin = 0,
    ymax = 0.5,
    xmin = 1,
    xmax = 10,
    xtick = data,
    ]
    \addplot[color=black, style=dashed] coordinates {(1, 0.152) (2, 0.208) (5, 0.314) (7,0.353) (10, 0.413)};
    \addplot[color=red, mark=x] coordinates {(1,0.125) (2,0.160) (5,0.246) (7,0.272) (10,0.306)};
    \addplot[color=green, mark=*] coordinates {(1,0.128) (2,0.175) (5,0.258) (7,0.291) (10,0.335)};
    \addplot[color=blue, mark=diamond] coordinates {(1,0.118) (2,0.168) (5,0.243) (7,0.279) (10,0.329)};
    \addplot[color=blue, mark=triangle] coordinates {(1,0.100) (2,0.132) (5,0.183) (7,0.204) (10,0.234)};
\end{axis}
\end{tikzpicture}

\begin{tikzpicture}[scale=0.55]
  \begin{axis}[
    title = {item similarity},
    legend columns = -1,
    xlabel={\textbf{k}},
    ymin = 0,
    ymax = 1,
    xmin = 1,
    xmax = 10,
    xtick = data,
    ]
    \addplot[color=black, style=dashed] coordinates {(1, 0.152) (2, 0.208) (5, 0.314) (7,0.353) (10, 0.413)};
    \addplot[color=red, mark=x] coordinates {(1,0.595) (2,0.725) (5,0.868) (7,0.898) (10,0.937)};
    \addplot[color=green, mark=*] coordinates {(1,0.132) (2,0.186) (5,0.283) (7,0.332) (10,0.384)};
    \addplot[color=blue, mark=diamond] coordinates {(1,0.233) (2,0.340) (5,0.486) (7,0.551) (10,0.605)};
    \addplot[color=blue, mark=triangle] coordinates {(1,0.602) (2,0.703) (5,0.805) (7,0.835) (10,0.860)};
\end{axis}
\end{tikzpicture}

\begin{tikzpicture}[scale=0.55]
  \begin{axis}[
    title = {item-tag similarity},
    legend columns = -1,
    legend entries = {global, $f_{mul}$, $f_{min}$, $f_{pow}$ ($\lambda=1.1$), $f_{pow}$ ($\lambda=2$)},
    legend to name = leg1,
    xlabel={\textbf{k}},
    ymin = 0,
    ymax = 1,
    xmin = 1,
    xmax = 10,
    xtick = data,
    ]
    \addplot[color=black, style=dashed] coordinates {(1, 0.152) (2, 0.208) (5, 0.314) (7,0.353) (10, 0.413)};
    \addplot[color=red, mark=x] coordinates {(0,0.707) (2,0.840) (5,0.933) (7,0.950) (10,0.966)};
    \addplot[color=green, mark=*] coordinates {(0,0.145) (2,0.201) (5,0.299) (7,0.344) (10,0.399)};
    \addplot[color=blue, mark=diamond] coordinates {(0,0.375) (2,0.499) (5,0.644) (7,0.706) (10,0.762)};
    \addplot[color=blue, mark=triangle] coordinates {(0,0.594) (2,0.693) (5,0.779) (7,0.806) (10,0.838)};
\end{axis}
\end{tikzpicture}
}
{\tiny \ref{leg1}}
\caption{Predicting unpopular bookmarks.\label{fig:relev_exp2}}
\end{figure*}
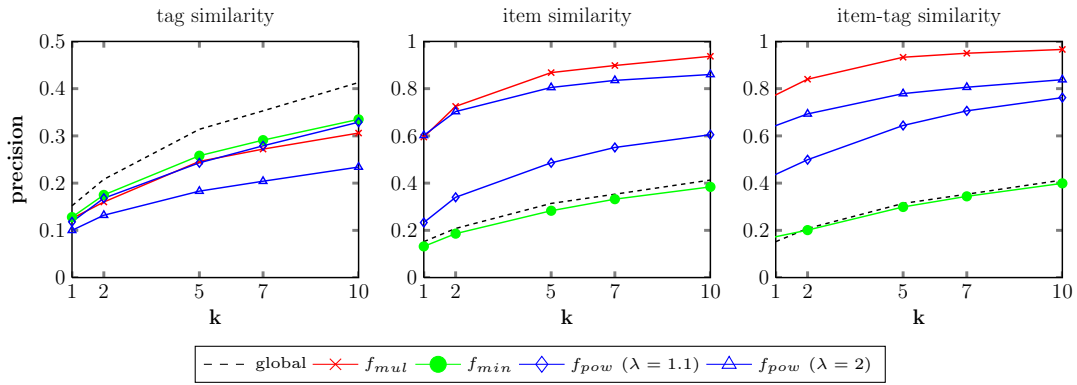

The results are presented in Figure~\ref{fig:relev_exp2}. They are similar to a good extent to those of the first experiment, with two main differences: (i) personalization fails completely in the tag similarity network, (ii) in the item and item-tag similarity networks personalization achieves considerably higher prediction performance than in the case of  predicting items tagged with popular tags. This is because, in the case of long-tail items, the functions that are skewed towards the closest users, i.e., $f_{mul}$ and $f_{pow},\lambda=2$, will rank higher the items belonging to the closest users. 

\begin{figure*}
\centering
\subfigure{
\begin{tikzpicture}[scale=0.55]
  \begin{axis}[
    title = {tag similarity},
    legend columns = -1,
    ylabel={\textbf{precision}},
    xlabel={\textbf{k}},
    ymin = 0,
    ymax = 0.5,
    xmin = 1,
    xmax = 10,
    xtick = data,
    ]
    \addplot[color=black, style=dashed] coordinates {(1, 0.035) (2, 0.067) (5, 0.128) (7,0.152) (10, 0.178)};
    \addplot[color=red, mark=x] coordinates {(1,0.057) (2,0.082) (5,0.132) (7,0.159) (10,0.189)};
    \addplot[color=green, mark=*] coordinates {(1,0.034) (2,0.069) (5,0.128) (7,0.143) (10,0.166)};
    \addplot[color=blue, mark=diamond] coordinates {(1,0.049) (2,0.076) (5,0.127) (7,0.153) (10,0.179)};
    \addplot[color=blue, mark=triangle] coordinates {(1,0.037) (2,0.056) (5,0.088) (7,0.101) (10,0.112)};
\end{axis}
\end{tikzpicture}

\begin{tikzpicture}[scale=0.55]
  \begin{axis}[
    title = {item similarity},
    legend columns = -1,
    xlabel={\textbf{k}},
    ymin = 0,
    ymax = 1,
    xmin = 1,
    xmax = 10,
    xtick = data,
    ]
    \addplot[color=black, style=dashed] coordinates {(1, 0.035) (2, 0.067) (5, 0.128) (7,0.152) (10, 0.178)};
    \addplot[color=red, mark=x] coordinates {(1,0.217) (2,0.408) (5,0.417) (7,0.559) (10,0.590)};
    \addplot[color=green, mark=*] coordinates {(1,0.031) (2,0.066) (5,0.118) (7,0.142) (10,0.166)};
    \addplot[color=blue, mark=diamond] coordinates {(1,0.095) (2,0.148) (5,0.222) (7,0.251) (10,0.288)};
    \addplot[color=blue, mark=triangle] coordinates {(1,0.301) (2,0.373) (5,0.465) (7,0.496) (10,0.517)};
\end{axis}
\end{tikzpicture}

\begin{tikzpicture}[scale=0.55]
  \begin{axis}[
    title = {item-tag similarity},
    legend columns = -1,
    legend entries = {global, $f_{mul}$, $f_{min}$, $f_{pow}$ ($\lambda=1.1$), $f_{pow}$ ($\lambda=2$)},
    legend to name = leg,
    xlabel={\textbf{k}},
    ymin = 0,
    ymax = 1,
    xmin = 1,
    xmax = 10,
    xtick = data,
    ]
    \addplot[color=black, style=dashed] coordinates {(1, 0.035) (2, 0.067) (5, 0.128) (7,0.152) (10, 0.178)};
    \addplot[color=red, mark=x] coordinates {(0,0.412) (2,0.501) (5,0.584) (7,0.611) (10,0.627)};
    \addplot[color=green, mark=*] coordinates {(0,0.036) (2,0.071) (5,0.131) (7,0.149) (10,0.177)};
    \addplot[color=blue, mark=diamond] coordinates {(0,0.185) (2,0.272) (5,0.357) (7,0.399) (10,0.426)};
    \addplot[color=blue, mark=triangle] coordinates {(0,0.332) (2,0.409) (5,0.477) (7,0.497) (10,0.515)};
\end{axis}
\end{tikzpicture}
}
{\tiny \ref{leg}}
\caption{Predicting bookmarks tagged with semi-popular tags.\label{fig:relev_exp1}}
\end{figure*}
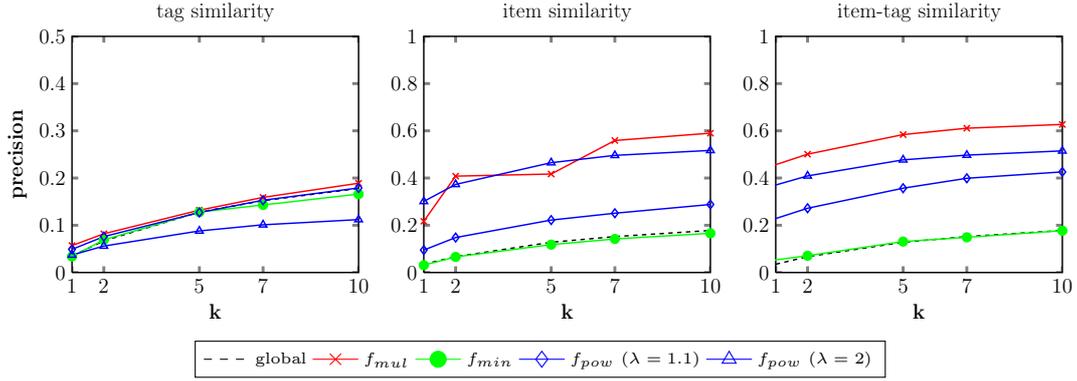

\section{Other Related Work\label{sec:related}}

The topic of search in a social setting has received increased attention lately. Studies and models of personalization of social tagging sites can be found in~\cite{Wang10,Heymann08,Dou07,Xu08}. Other studies have found that including social knowledge in scoring models can improve search and recommendation algorithms. In~\cite{Carmel09}, personalization based on a similarity network is shown to outperform other personalization approaches and the non-personalized social search. A study on a last.fm dataset in~\cite{Konstas09} has found that incorporating social knowledge in a graph model system improves the retrieval recall of music track recommendation algorithms.  An architecture  for social data management is given in~\cite{Amer-Yahia09,Amer-Yahia09SS}, along with a framework for information discovery and presentation in social content sites.   Another approach to rank resources in social tagging environments is CubeLSI~\cite{Bi11}, which uses a vector space model and extends Latent Semantic Indexing to include taggers in the feature space of resources, in order to better match queries to documents. 
 FolkRank~\cite{Hotho06} proposes a ranking model in social bookmarking sites, for recommendation and search, based on an adaptation of PageRank over the tripartite graph of users, tags and resources. It follows the intuition that a resource that is tagged with important tags by important users becomes important itself and, symmetrically, for tags and users. An  alternative approach to social-aware search, using personalized PageRank, was presented in~\cite{Bao07}. There, the same tripartite model of annotators, resources and annotations is used to compute measures of similarities between resources and queries, and  to capture the social popularity of  resources. However, none of these approaches incorporate the user-to-user relationships in their ranking model. In contrast, the social network is an integral part of the scoring model in our setting, if not the decisive one, while this network can have various semantics (e.g., tagging similarity, activity similarity or even trust). 
 
 The scoring model used in~\cite{Schenkel08} is revisited in~\cite{Yin10}. There, a textual relevance and a social influence score are combined in the overall scoring of items, the latter being computed as the inverse of the shortest path between the seeker and the document publishers. This  model is also used in the context of top-$k$ retrieval of spatial web objects~\cite{Cao10}, where a prestige-based relevance score is computed by combining the overall relevance of an object with its spatial distance. 

\section{Conclusions and Future Work}\label{sec:future}
We considered in this paper top-k query answering in social bookmarking applications,  proposing algorithms that have the potential to scale in real applications, in an online context where the social network,  the tagging data and even the seekers' search ingredients can change at any moment. Our solutions address the main drawbacks of previous approaches. With respect to applicability and scalability, we avoid expensive and hardly updatable pre-computations of proximity values, by an on-the-fly approach. We show that it is applicable to a wide family of functions for proximity computation in a social network. With respect to efficiency, we show that $\textsc{TOPKS}$ is instance optimal in the exclusively social context and, via extensive experiments, that it performs significantly better than the algorithm from previous literature. We also considered widely-applicable  approximate techniques, showing  they have the potential to drastically reduce computation costs, while exhibiting high accuracy. 

We see many directions for future work. As mentioned in the previous section, optimizing the branch choice heuristic is a promising direction that we plan to explore further. Experimenting with other aggregation functions,  probabilistic bounds using statistics  tailored to certain assumptions (e.g,  for power-law distributions) or richer descriptions for proximity vectors and term-frequencies are other important directions. We are also investigating approaches for computing  results in a distributed style, when one has access to query results pertaining to various seekers, or when the same query is run at various points in the network. Finally, we intend to adapt our approach to deal with networks containing also negative links (e.g., trust / distrust networks).

\bibliographystyle{abbrv} 
\bibliography{main}

\appendix

\section{Proof of Theorem 1}\label{sec:proof}

\begin{proof}
Since on each access to a user list, all items tagged by the respective user with any of the query terms are retrieved, the position in the proximity vector at any step in the run of the algorithm  is not tag-dependent. So $cost({\cal A,D})$ is equal to the position $p$ in the seeker's proximity vector at the moment of $\cal{A}$'s termination. Throughout the proof, we use the subscript $p$ to  denote the value of a given variable at step $p$ in the execution of $\cal{A}$.
  We will use a proof argument similar in style to the one for NRA~\cite{Fagin01}.

Let us assume that $\textsc{TOPKS}_{\alpha=0}$ does not stop at position $p$ (in the proximity vector) and that there exists an algorithm ${\cal A}\neq \textsc{TOPKS}_{\alpha=0}$ that does. 
	
Since $\textsc{TOPKS}_{\alpha=0}$ does not stop at position $p$,  there exists an item $r \not \in \{D_{p}[1],\dots,D_{p}[k]\}$ having
$\textsc{MaxScore}_{p}(r,Q)> \textsc{MinScore}_{p}(D_{p}[k],Q),$ and
$\textsc{MinScore}_{p}(r,Q)\leqslant$\\ $\textsc{MinScore}_{p}(D_{p}[i],Q),\forall i\in \{1,\dots,k\}.$
 If $\textsc{MinScore}_{p}(r,Q)=\textsc{MinScore}_{p}(D_p[k],Q)$ then necessarily  $\textsc{MaxScore}_{p}(r,Q)\leqslant\textsc{MaxScore}_{p}(D_p[k],Q)$ (ties for pessimistic scores are broken by the optimistic ones, then arbitrarily for the optimistic scores). 
	
	In ${\cal D}$,  we assume that at step $p$  we have  with $\textsc{TOPKS}_{\alpha=0}$ in the current (unconsumed) position in each of the $|Q|$ inverted lists $IL(t_j)$ an item  $v_j$, necessarily not yet candidate. By definition, for any algorithm  ${\cal A}\in \textbf{S}$, for any tag $t_j$ of the input $Q$,   ${\cal A}$ is \emph{at most as advanced} in the inverted list $IL(t_j)$ as $\textsc{TOPKS}_{\alpha=0}$. Without loss of generality, let us assume ${\cal A}$ is as advanced as $\textsc{TOPKS}_{\alpha=0}$.

Towards a contradiction, showing that ${\cal A}$ is not sound over all possible inputs,  we  will construct an instance ${\cal D}'$, which is equal to ${\cal D}$ up to position $p$.  We consider the following two possible cases:

\emph{Case 1:  ${\cal A}$ outputs $r$ as one of the top-$k$ item,} 
i.e., there do not exist $k$ items having a higher score than $r$.  

In ${\cal D}'$ will start from what ${\cal A}$ could have already read and used, including the items $v_j=top\_item_p(t_j)$ and the value $top_p(H)$ (the proximity value of the $p+1$ user).

 ${\cal D}'$  will  be such that $\textsc{Score}(r,Q)=\textsc{MinScore}_{p}(r,Q)$, and $textsc{Score}(D_p[i],Q)=\textsc{MaxScore}_{p}(D_p[i],Q), \forall i\in\{1,\dots,k\}.$
Now, for each $D_p[i]$, if $\textsc{MaxScore}_p(D_p[i],Q)>$\\$ \textsc{MinScore}_p(D_p[i],Q)$, i.e.,  we do not have $D_p[i]$'s final score at step $p$, we assume the following in ${\cal D}'$. For each $t_j\in Q$ for which $tf(D_p[i],t_j)$ is unknown, we assume that we have $D_p[i]$ in $IL(t_j)$ after $v_j$, with $tf(D_p[i],t_j)=top\_tf_{p}(t_j)$.  
 Also,  for every $t_j\in Q$ we set in the proximity vector, after $p+1$, the next $x_{ij}=unseen\_users(D_p[i],t_j)$ values to $top_p(H)$, making also $D_p[i]$ present  in each of these users' lists for $t_j$. 
 By doing so,  the exact score of each $D_p[i]$, $i\in\{1,\dots,k\}$, is equal to the maximal possible one at step $p$; 
after $max_{i,j}(x_{ij})$ steps, all these $k$ scores $\textsc{Score}(D_p[i],Q)$ would be computed. 

For item $r$, for each $t_j\in Q$ for which we do not have $tf(r,t_j)$, since  $r$ must come later in $IL(t_j)$ (after $v_j$), we can assume that  $tf(r,t_j)=partial\_tf(r,t_j)$ (this makes\\ $unseen\_users(r,t_j)=0$). Also, for every $t_j\in Q$ for which we do know $tf(r,t_j)$, after the required $max_{i,j}(x_{ij})$ proximity values set as described previously, we set the next $unseen\_users(r,t_j)$ in the proximity vector to $0$, with each of these users having tagged $r$ with $t_j$.   All this  ensures that $\textsc{MinScore}_p(r,Q)=\textsc{Score}(r,Q)$. 

We can now contradict the correctness of algorithm ${\cal A}$,  showing  that $\textsc{Score}(r,Q)<$\\ $\textsc{Score}(D_p[i],Q)$ for all  $i$.  

We have the following inequalities:

\begin{eqnarray}
\textsc{MinScore}_p(D_p[k],Q)  \geqslant	 \textsc{MinScore}_{p}(r,Q)~~	\label{equ:tk1}\\
	\textsc{MinScore}_p(D_p[k],Q)\leqslant  \textsc{MinScore}_p(D_p[i],Q), \forall i ~~ \label{equ:tk2}\\
	\textsc{MinScore}_p(D_p[i],Q)  \leqslant  \textsc{MaxScore}_p(D_p[i],Q), \forall i~~ 
	\label{equ:tk3}
\end{eqnarray}

If $\textsc{MinScore}_p(r,Q)<\textsc{MinScore}_p(D_p[k],Q)$ then it follows from Eq.~(\ref{equ:tk1}),~(\ref{equ:tk2}),~(\ref{equ:tk3}) that 
$$\textsc{Score}(r,Q)<\textsc{Score}(D_p[i],Q), \forall i.$$

If $\textsc{MinScore}_p(r,Q)=\textsc{MinScore}_p(D_p[k],Q)$ then, for each $i\in \{1,\dots,k\}$, if:
\begin{enumerate}
\item $\textsc{MinScore}_p(D_p[k],Q)=\textsc{MinScore}_p(D_p[i],Q)$:  we have $\textsc{MaxScore}_p(r,Q)>$\\ $\textsc{MinScore}_p(D_p[k],Q)$ and $\textsc{MaxScore}_p(r,Q)\leqslant \textsc{MaxScore}_p(D_p[i],Q)$;  it follows that $\textsc{Score}(r,Q)<$ $\textsc{Score}(D_p[i],Q)$,
\item $\textsc{MinScore}_p(r,Q)<\textsc{MinScore}_p(D_p[k],Q)$:  we have $\textsc{MinScore}_p(D_p[k],Q)<$\\ $\textsc{Score}(D_p[i],Q)$; it follows that $\textsc{Score}(r,Q)<\textsc{Score}(D_p[i],Q)$.
\end{enumerate}
Hence, in any possible configuration,  $r$ is not in the top-$k$ result over ${\cal D}'$. But since ${\cal D}'$ and ${\cal D}$ are indistinguishable by algorithm ${\cal A}$, which stops at step $p$ outputting $r$ in the result, this contradicts ${\cal A}$'s correctness.

\eat{ 
Taking into account the new items $v_j\not\in D_p$ we have the following two cases: (i) there exist $v_j$ that could have entered the top-$k$, but ${\cal A}$ was unable to know about them at its termination, and in that case ${\cal A}$ is mistaken by not outputting $v_j$ also into the top-$k$, and (ii) no $v_j$ could have entered the top-$k$, and in this case ${\cal A}$ is mistaken by outputting $r$ as a top-$k$ item.
}


	
\emph{Case 2:  ${\cal A}$ does not output $r$ as a top-$k$ item}, which means that ${\cal A}$ assumes that the final score of $r$, $\textsc{Score}(r,Q)$ is not in the top-$k$ scores for ${\cal D}$.   

${\cal D}'$, undistinguishable from ${\cal D}$ up to position $p$,   will now be such that $\textsc{Score}(r,Q)=$\\ $\textsc{MaxScore}_{p}(r,Q)$ and $\textsc{Score}(D_p[i],Q)=\textsc{MinScore}_{p}(D_p[i],Q)$, for each $D_p[i] \in D_p$ s.t.  $D_p[i]\not=r$.


If $r$'s score at step $p$ is not already the final one, i.e., $\textsc{MaxScore}_p(r,Q)=\textsc{MinScore}_p(r,Q)$, we assume the following in ${\cal D}'$:
 for each tag $t_j\in Q$ for which $tf(r,t_j)$ is yet unknown, we assume that $r$ comes later (after $v_j$) in  $IL(t_j)$,  having $tf(r,t_j)=top\_tf_{p}(t_j)$. Then, for every $t_j\in Q$ we set in the proximity vector, after the $p+1$ position, the next 
 $x_{j}=unseen\_users(r,t_j)$ values to $top_p(H)$,  making also $r$ present  in each of these users' lists for $t_j$.  
 
By this, the exact score of $r$ is equal to the maximal possible one at step $p$; 
after $max_{j}(x_{j})$ steps, the score $\textsc{Score}r,Q)$ would be computed. 

Symmetrically, for each each $D_p[i] \in D_p$ s.t. $D_p[i]\not=r$,  and each $t_j\in Q$ for which $tf(D_p[i],t_j)$ is yet unknown, we assume that $D_p[i]$ comes later (after $v_j$) in $IL(t_j)$,  having $tf(D_p[i],t_j)=partial\_tf(D_p[i],t_j)$ (hence $unseen\_users(D_p[i],t_j)=0$). Then, for every $t_j\in Q$ for which we know $tf(D_p[i],t_j)$, after the $max_{j}(x_{j})$ values set as described previously in the seeker's proximity vector, we set the next $y_{ij}=unseen\_users(D_p[i],t_j)$ values  to $0$, making also $D_p[i]$ present  in each of these users' lists for $t_j$. This construction ensures that, the exact score of each  $D_p[i]$ is equal to the minimal possible one at step $p$; 
after $max_{i,j}(y_{ij})$ steps, all these scores $\textsc{Score}(D_p[i],Q)$ would be computed.

Since we have that
\vspace{-2mm}
 $$\textsc{Score}(r,Q)=\textsc{MaxScore}_p(r,Q)>\textsc{MinScore}_p(D_p[k],Q)$$ and  $\textsc{MinScore}_p(D_p[k],Q)=\textsc{Score}(D_p[k],Q)$,  given that  for every item  $D_p[l]$, $l>k$ s.t. $D_p[l]\not=r$ we have $\textsc{Score}(D_p[l],Q)\leqslant \textsc{MinScore}_p(D_p[k],Q)$,  $r$ should be among the top-$k$ items in ${\cal D}'$. But since ${\cal D}'$ and ${\cal D}$ are indistinguishable by algorithm ${\cal A}$, which stops at step $p$ without outputting $r$ in the result, this contradicts 
${\cal A}$'s correctness.

In this proof, we have ignored $\textsc{MaxScoreUnseen}(Q)$ in the inequalities. The unseen items can be simulated by adding  one virtual item $i_v$ to $D$, which does not exist and will never be encountered in user lists, with  $\textsc{MinScore}(i_v,Q)=0$ and\\ $\textsc{MaxScore}(i_v,Q)=\textsc{MaxScoreUnseen}(Q)$.  Then, the same proof argument applies to these items. 
\end{proof}

\section{Other $\sigma^+$ functions}

We present experimental results for the $f_{pow}$, $\lambda=1.1$, in Figures~\ref{fig:tagsim_pow},~\ref{fig:docsim_pow} and~\ref{fig:doctagsim_pow}). While the results follow the same trend as those of $f_{mul}$, one can notice that the speedups achieved by the \textsc{TOPKS} variants are directly affected by the speed of the ``drop'' in proximity values. Generally, $f_{mul}$ values drop faster than those of $f_{pow}$. 

\begin{figure*}\small
\centering
\subfigure[BM15]{
\subfigure[$\alpha=0$]{

}
}
\caption{Abstract cost and running time comparison over the  tag-similarity network and the $f_{pow}$, $\lambda=1.1$ proximity function (red: top-10, yellow: top-20).\label{fig:tagsim_pow}}
\end{figure*}

\begin{figure}\small
\centering
\subfigure[BM15]{
\subfigure[$\alpha=0$]{
%
}
}
\caption{Abstract cost and running time comparison over the  item-similarity network and the $f_{pow}$, $\lambda=1.1$ proximity function (red: top-10, yellow: top-20).\label{fig:docsim_pow}}
\end{figure}

\begin{figure}
\centering
\subfigure[BM15]{
\subfigure[$\alpha=0$]{
%
}
}
\caption{Abstract cost and running time comparison over the  item-tag similarity network and the $f_{pow}$, $\lambda=1.1$, proximity function (red: top-10, yellow: top-20).\label{fig:doctagsim_pow}}
\end{figure}

\eat{
\begin{figure*}\small
\centering
\subfigure[BM15]{
\subfigure[$\alpha=0$]{
%
}
}
\caption{Abstract cost and running time comparison over the  tag-similarity network and the $f_{min}$ proximity function (red: top-10, yellow: top-20).\label{fig:tagsim_min}}
\end{figure*}

\begin{figure}\small
\centering
\subfigure[BM15]{
\subfigure[$\alpha=0$]{
%
}
}
\caption{Abstract cost and running time comparison over the  item-similarity network and the $f_{min}$ proximity function (red: top-10, yellow: top-20).\label{fig:docsim_min}}
\end{figure}

\begin{figure}
\centering
\subfigure[BM15]{
\subfigure[$\alpha=0$]{
%
}
}
\caption{Abstract cost and running time comparison over the  item-tag similarity network and the $f_{min}$ proximity function (red: top-10, yellow: top-20).\label{fig:doctagsim_min}}

\end{figure}

}

\end{document}